\pdfoutput=1

\documentclass[fleqn,usenatbib]{mnras}

\usepackage{newtxtext,newtxmath}
\usepackage[T1]{fontenc}

\DeclareRobustCommand{\VAN}[3]{#2}
\let\VANthebibliography\thebibliography
\def\thebibliography{\DeclareRobustCommand{\VAN}[3]{##3}\VANthebibliography}

\usepackage{graphicx}	% Including figure files
\usepackage{amsmath}		% Advanced maths commands
\title[Heliospheric ENAs: modeling and comp. with {\it IBEX-Hi}]{Heliospheric Energetic Neutral Atoms: Non-stationary Modeling and Comparison with {\it IBEX-Hi} data}

\author[I. I. Baliukin et al.]{
I. I. Baliukin,$^{1,2,3}$\thanks{E-mail: igor.baliukin@gmail.com}
V. V. Izmodenov,$^{1,2,4}$
and D. B. Alexashov$^{4}$
\\
$^{1}$Space Research Institute of Russian Academy of Sciences, Profsoyuznaya Str. 84/32, Moscow, 117335, Russia\\
$^{2}$Moscow Center for Fundamental and Applied Mathematics, Lomonosov Moscow State University, GSP-1, Leninskie Gory, Moscow, 119991, Russia\\
$^{3}$National Research University Higher School of Economics, Moscow, Russia\\
$^{4}$Institute for Problems in Mechanics, Vernadskogo 101-1, Moscow, 119526, Russia
}

\date{Accepted XXX. Received YYY; in original form ZZZ}

\pubyear{2020}

\begin{document}
\label{firstpage}
\pagerange{\pageref{firstpage}--\pageref{lastpage}}
\maketitle

\begin{abstract}

The Interstellar Boundary Explorer ({\it IBEX}) has been measuring fluxes of the Energetic Neutral Atoms (ENAs) using the {\it IBEX-Hi} (0.3 -- 6 keV) instrument since 2008. We have developed a numerical time-depended code to calculate globally distributed flux (GDF) of hydrogen ENAs employing both 1) 3D kinetic-MHD model of the global heliosphere and 2) reconstruction of atom trajectories from 1 au, where they are observed by {\it IBEX}, to the point of their origin in the inner heliosheath (IHS). The key factor in the simulation is a detailed kinetic consideration of the pickup ions (PUIs), the supra-thermal component of protons in the heliosphere, which is ``parental" to the ENAs and originates in the region of the supersonic solar wind being picked by the heliospheric magnetic field.
As a result of our study, we have concluded that
(1) the developed model is able to reproduce the geometry of the multi-lobe structure seen in the {\it IBEX-Hi} GDF maps,
(2) the GDF is extremely sensitive to the form of the velocity distribution function of PUIs in the IHS, and the accounting for the existence of an additional energetic population of PUIs is essential to explain the data,
(3) despite a relatively good agreement, there are some quantitative differences between the model calculations and {\it IBEX-Hi} data. Possible reasons for these differences are discussed.

\end{abstract}


%%%%%%%%%%%%%%%%%%%%%%%%%%%%%%%%%%%%%%%%%%%%%%%%%%

%%%%%%%%%%%%%%%%% BODY OF PAPER %%%%%%%%%%%%%%%%%%

\section{Introduction} \label{sec:intro}

The solar system is surrounded by the local interstellar medium (LISM) and moving through it with a bulk velocity of $\sim$26 km s$^{-1}$ \citep[e.g.,][]{witte2004, mccomas2015}. The interaction of the solar wind (SW) with the ionized component of the LISM forms a complex structure that is called the heliospheric interface. The heliopause (HP) is a tangential discontinuity that separates the SW and interstellar plasmas from each other. There are two shocks in the heliospheric interface -- the termination shock (TS), where the SW is slowed down from supersonic to subsonic speed, and a bow shock (BS), where the interstellar flow is slowed down, but the existence of latter is under discussion \citep[see, e.g.,][]{izmod2009,mccomas2012,zank2013}. The region of the compressed and heated plasma between the shocks is commonly called a heliosheath and the region between the TS and HP -- the inner heliosheath (IHS).

The neutral component of the LISM consists mainly of hydrogen atoms. The interstellar H atoms have a large mean free path for charge exchange, which is comparable with the characteristic size of the heliosphere \citep{izmod2001}, and due to the relative motion of the Sun and LISM, they can penetrate the heliosphere. In the heliosheath, H atoms may experience charge exchange with hot protons, which results in the production of energetic neutral atoms (ENAs). Some of these ENAs propagate close to the Sun and Earth's orbit, where they can be measured.

The {\it Interstellar Boundary Explorer} ({\it IBEX}) spacecraft was launched into a highly elliptical orbit around Earth in October 2008 to obtain the first all-sky maps of the neutral gas/plasma interactions at the heliospheric boundary and to directly sample the interstellar gas flow through the inner heliosphere \citep{mccomas2009}. To achieve these goals {\it IBEX} is a Sun-pointed spinning satellite, which carries two energetic atom sensors: {\it IBEX-Lo} (0.01 -- 2 keV) and {\it IBEX-Hi} (0.3 -- 6 keV). A detailed description of the {\it IBEX-Hi} sensor may be found in \citet{funsten2009}. The {\it IBEX-Hi} instrument is measuring ENA fluxes and these data are one of the few sources of knowledge about the structure of the heliospheric boundary, imposing significant limitations on the parameters of the heliospheric models. \citet{mccomas2020} have examined IBEX’s global ENA observations over a full solar activity cycle (Solar Cycle 24), covering 11 years from 2009 through 2019.

The observations of ENAs by {\it IBEX-Hi} have revealed two populations: one of them is emitted from a narrow circular part of the sky that is called ``ribbon", and a globally distributed flux (GDF) that is controlled by processes in the heliosheath \citep{mccomas2009}. \citet{schwadron2011,schwadron2014} have performed the analysis of {\it IBEX-Hi} data and developed the technique to separate ENA emissions in the ribbon from the GDF, which, as it turned out, has a complex multi-lobe structure. It is commonly assumed that the GDF originates from the IHS, where the supersonic SW and pickup ions (PUIs) are slowed and heated after crossing the TS. The PUIs are formed in a result of ionization of interstellar hydrogen atoms mainly through charge exchange in the heliosphere, where they are picked by the heliospheric magnetic field. The ribbon, in turn, is formed by secondary charge exchange in the outer heliosheath \citep[see, e.g., ][]{mccomas2009,chalov2010,heerikhuisen2010}.

There are two populations of protons in the heliosphere -- the cold thermal population of core SW protons and the hot supra-thermal population of PUIs. The PUIs in the inner heliosheath can be divided into subpopulations, transmitted or reflected, depending on its interaction with TS \citep{zank2010}. In some works on the modeling of ENA fluxes the authors made attempts to model the quite distinct populations – thermal and pickup protons – using one kappa-distribution \citep[e.g.,][]{heerikhuisen2008} or a superposition of Maxwell distributions \citep{zank2010,zirnstein2017,kornbleuth2018,shrestha2020}.

In this paper, we investigate the GDF maps produced in the frame of the latest heliospheric model of the Moscow group \citep{izmod2015,izmod2020}, and perform its comparison with {\it IBEX-Hi} data \citep{schwadron2014}. The key factor in the simulation is a detailed kinetic consideration of the PUIs, the supra-thermal component of protons in the heliosphere. The paper is organized as follows. Section \ref{sec:fluxes} describes the method of calculation of the ENA fluxes. In Section \ref{sec:model}, the detailed description of the PUI distribution model is provided. Section \ref{sec:partitioning} describes the technique of partitioning charged particles into components from single-fluid plasma calculations. In Section \ref{sec:results}, the results of the numerical calculations and their comparison with {\it IBEX-Hi} data are presented. In Section \ref{sec:tail}, the qualitative effect of additional energetic PUI population on the ENA fluxes is discussed. Finally, Section \ref{sec:conclusions} provides an overall summary of our work.

\section{Modeling of the ENA fluxes} \label{sec:fluxes}

The primary ENAs, which are the source of GDF, are born in charge exchange between the SW protons and H atoms in the inner heliosheath (IHS). The directional differential ENA flux (in the solar inertial reference frame) is a line of sight (LOS) integral:
\begin{equation}
j_{\rm ENA}(t_{\rm obs}, \mathbf{r}_{\rm obs}, v, \mathbf{LOS}) = \frac{1}{m_{\rm H}} \int_{s_{\rm TS}}^{s_{\rm HP}} \nu_{\rm H}(t, \mathbf{r},\mathbf{v}) f(t, \mathbf{r}, \mathbf{v}) v S_{\rm p,ENA} ds,\:
\label{eq:flux}
\end{equation}
where $t_{\rm obs}$ is the moment of observation, $\mathbf{r}_{\rm obs}$ is the position of the observer, $\mathbf{LOS}$ is the unit vector in the line of sight direction, $\mathbf{v} = v \cdot \mathbf{LOS}$ is the velocity of an ENA, $m_{\rm H}$ is the mass of H atom, $f = f_{\rm sw} + f_{\rm pui}$ is the velocity distribution of protons in the IHS (the sum of core SW proton and PUI distribution functions). The integration is performed along the ENA trajectory in the IHS (in the region between the TS and HP), with $ds = v dt$ being the differential path length. The values of the variable $s$ along the trajectory $s_{\rm TS}$ and $s_{\rm HP}$ correspond to atom intersections of the TS and HP, respectively. In principle, the velocity of an ENA is changing along the trajectory due to the influence of gravitational and radiation pressure forces by the Sun, i.e. $v = v(s)$, but for the energies under consideration (in the {\it IBEX-Hi} energy range) the change of velocity is negligible, so we assume that the velocity is constant along the trajectory.

In Equation (\ref{eq:flux}) $\nu_{H}$ is the production rate of ENAs due to the charge exchange of protons with H atoms, which is defined as
\begin{equation}
\nu_{\rm H} (t, \mathbf{r}, \mathbf{v}) =  \int\int\int f_{\rm H}(t, \mathbf{r}, \mathbf{v}_{\rm H}) |\mathbf{v} - \mathbf{v}_{\rm H}| \sigma_{\rm ex}(|\mathbf{v} - \mathbf{v}_{\rm H}|) \, d\mathbf{v}_{\rm H},
\label{eq:nuH}
\end{equation}
where $f_{\rm H}(t, \mathbf{r}, \mathbf{v}_{\rm H})$ is the H velocity distribution function and $\sigma_{\rm ex}$ is the effective charge exchange cross section that depends on the relative atom-proton velocity. In our calculations, the cross section from \citet{lindsay2005} was used. The extinction of ENAs is determined by survival probability
$$
S_{\rm p,ENA} = \exp{ \left(- \int_{t}^{t_{\rm obs}} \nu_{\rm ion}(\tau, \mathbf{r}(\tau), \mathbf{v}(\tau)) d\tau \right)}, 
$$
where $\nu_{ion}$ is the total ionization rate due to the ionization processes (charge exchange with protons, photoionization, and electron impact), i.e. $\nu_{\rm ion} = \nu_{\rm ex} + \nu_{\rm ph} + \nu_{\rm imp}$. In our calculations we neglect electron impact ionization ($\nu_{\rm imp} = 0$), and assume that $\nu_{\rm ph} = \nu_{\rm ph,E}(t, \lambda) \cdot (r_{\rm E} / r)^2$, where $\lambda$ is heliolatitude and $r_{\rm E}$ = 1 au. The temporal and heliolatitudinal variations of the photoionization rate $\nu_{\rm ph,E}(t, \lambda)$ at 1 au adopted in our model were obtained from different experimental data ({\it OMNI}, {\it SWAN/SOHO}) the same way as it was performed in \citet{katushkina2015}. For the stationary model calculations we assume constant photoionization rate at 1 au $\nu_{\rm ph,E} = 1.67 \times 10^{-7}\:s^{-1}$ as it was taken in \citet{izmod2015}. The charge exchange ionization rate $\nu_{\rm ex}$ is calculated as
$$
\nu_{\rm ex}(t, \mathbf{r}, \mathbf{v}) = \int\int\int f(t, \mathbf{r}, \mathbf{v}_{p}) |\mathbf{v}-\mathbf{v}_{\rm p}| \sigma_{\rm ex}(|\mathbf{v}-\mathbf{v}_{\rm p}|)\, d\mathbf{v}_{\rm p},
$$
where $\mathbf{v}_{\rm p}$ is the velocity of proton.

To calculate the differential fluxes measured by {\it IBEX-Hi} at different energy channels, the energy transmission of {\it IBEX-Hi} electrostatic analyzers were taken into account (for details, see Appendix \ref{app:flux_ibex-hi}). Accounting for energy transmission leads to the re-distribution of fluxes between the energy channels and the spreading of the observed spectrum. Important to note, that it is possible to make some simplifications in order to optimize the calculations. For energies under consideration the production rate of ENAs $\nu_{\rm H}$ can be safely approximated in Equation \ref{eq:nuH} as $\nu_{\rm H} (t, \mathbf{r}, \mathbf{v}) \approx n_{\rm H}(t, \mathbf{r}) v \sigma_{\rm ex}(v)$, since $|\mathbf{v} - \mathbf{v}_{\rm H}| \approx v$, so we use this simplification in our simulations.

Thus, to calculate the fluxes of ENAs, we need to know the velocity distribution function of protons in the heliosphere. The following section will describe the method to calculate the distribution of PUIs using the plasma and interstellar H atom distributions in the heliosphere obtained in the frame of the global kinetic-MHD model by \citet{izmod2020}.

\section{Model of PUI distribution} \label{sec:model}
 
\subsection{Kinetic model}
%When the deviation from isotropy is small, the diffusion equation for the pitch-angle-averaged distribution function is used:

The distribution function of PUIs is anisotropic in the case of weak scattering, i.e. when the SW turbulence level is low (the quiet solar wind). However, as follows from theoretical estimates and observations \citep{gloeckler1994}, the distribution function is almost isotropic in the disturbed SW, so the process of an effective pitch-angle diffusion must be operative. Therefore, we assume that the velocity distribution of pickup protons in the SW rest frame is isotropic, and it is determined by the velocity distribution function $f_{\rm pui}(t, \mathbf{r}, \mathbf{v})$ in the heliocentric coordinate system by the expression:
$$
f_{\rm pui}^*(t, \mathbf{r}, w) = \frac{1}{4 \pi}\int\int f_{\rm pui}(t, \mathbf{r}, \mathbf{v}) \sin\theta d \theta d \varphi,
$$
where $\mathbf{v} = \mathbf{V}(\mathbf{r},t) + \mathbf{w}$, $\mathbf{v}$ and $\mathbf{V}$ are velocity of pickup proton and bulk velocity of the plasma in the heliocentric coordinate system, $\mathbf{w}$ is the velocity of the pickup proton in the SW rest frame, and ($w$, $\theta$, $\varphi$) are coordinates of $\mathbf{w}$ in the spherical coordinate system. The kinetic equation for $f_{\rm pui}^*(t, \mathbf{r}, w)$ can be written in the following general form \citep[see, e.g.][]{isenberg1987,chalov2003}:
\begin{equation}
\frac{\partial f_{\rm pui}^*}{\partial t} + \mathbf{V} \cdot \frac{\partial f_{\rm pui}^*}{\partial \mathbf{r}} =
\frac{1}{w^2}\frac{\partial}{\partial w}\left( w^2 D \frac{\partial f_{\rm pui}^*}{\partial w} \right) +
\frac{w}{3} \frac{\partial f_{\rm pui}^*}{\partial w} {\rm div}(\mathbf{V})+S(t, \mathbf{r},w),
\label{eq:fpui_kinetic}
\end{equation}
taking into account velocity diffusion (where $D(t, \mathbf{r}, w) $ is the velocity diffusion coefficient) but ignoring spatial diffusion.
The estimations of the spatial diffusion coefficient \citep[e.g.,][its table 4]{scherer1998} show that for the energies under consideration the spatial diffusion can be neglected \citep[see, also,][]{rucinski1993, chalov1997}.
The source term $S(t, \mathbf{r}, w) = S_+(t, \mathbf{r}, w)- f^{*}_{\rm pui}(t, \mathbf{r}, w) S_-(t, \mathbf{r}, w)$, where $S_+$ and $S_-$ are responsible for production and losses (extinction) of PUIs, respectively, and can be calculated as
\begin{equation}
S_+(t, \mathbf{r}, w) = \frac{1}{4 \pi} \int \int f_{\rm H}(t, \mathbf{r}, \mathbf{V} + \mathbf{w}) \nu_{\rm ion}(t, \mathbf{r}, \mathbf{V} + \mathbf{w}) \sin \theta d \theta d \varphi,
\label{eq:s1term}
\end{equation}
\begin{equation}
S_-(t, \mathbf{r}, w) = \frac{1}{4 \pi} \int \int \nu_{\rm H}(t, \mathbf{r}, \mathbf{V} + \mathbf{w}) \sin \theta d \theta d \varphi.
\label{eq:s2term}
\end{equation}

In the IHS, neutrals can interact both with SW protons and PUIs through the charge exchange process, so $\nu_{\rm ex} = \nu_{\rm ex,sw} + \nu_{\rm ex,pui}$ and
\begin{equation}
\begin{gathered}
\nu_{\rm ex,sw}(t, \mathbf{r}, \mathbf{V} + \mathbf{w}) = \int\int\int f_{\rm sw}(t, \mathbf{r}, \mathbf{V} + \mathbf{w}_{\rm sw}) |\mathbf{w}-\mathbf{w}_{\rm sw}| \cdot \\
\cdot \sigma_{\rm ex}(|\mathbf{w}-\mathbf{w}_{\rm sw}|)\, d\mathbf{w}_{\rm sw},
\end{gathered}
\label{eq:nu_ex_sw}
\end{equation}

\begin{equation}
\begin{gathered}
\nu_{\rm ex,pui}(t, \mathbf{r}, \mathbf{w}) = \int f_{\rm pui}^*(t, \mathbf{r}, w_{\rm pui}) \cdot \\
\cdot \left( \int\int |\mathbf{w}-\mathbf{w}_{\rm pui}| \sigma_{\rm ex}(|\mathbf{w}-\mathbf{w}_{\rm pui}|) w_{\rm pui}^2 \sin \theta d\theta d\varphi \right) dw_{\rm pui}.
\end{gathered}
\label{eq:nu_ex_pui}
\end{equation}

\subsection{Method of characteristics}
In this paper, we consider a simple model and adopt $D = 0$ corresponding to a quiet SW, when the magnetic field fluctuation level is low \citep{chalov2003}, i.e. we neglect the velocity diffusion. Nevertheless, we admit that the process of velocity (energy) diffusion is connected with effective pitch-angle diffusion, and therefore should be taken into account. The study of this aspect will be held in future works. 

In this case, Equation (\ref{eq:fpui_kinetic}) becomes the first-order linear differential equation that can be solved by the method of characteristics. The characteristic is the SW particle trajectory
\begin{equation}
\frac{d\mathbf{r}}{dt} = \mathbf{V},
\label{eq:trajectory}
\end{equation}
which for the stationary case is also a streamline, and PUI velocity changes along it according to
\begin{equation}
\frac{d w}{dt} = -\frac{w}{3} {\rm div}(\mathbf{V}).
\label{eq:fpui_w}
\end{equation}

Equation (\ref{eq:fpui_kinetic}) has the following solution
\begin{equation}
\begin{split}
f_{\rm pui}^{*}(t, \mathbf{r}(t), w(t)) 
&= \int^t_{t_0} S_+(\tau, \mathbf{r}(\tau), w(\tau)) S_{\rm p,pui}(\tau, t) d\tau + \\
&+ f_{\rm pui}^{*}(t_0, \mathbf{r}(t_0), w(t_0)) S_{\rm p,pui}(t_0, t),
\end{split}
\label{eq:solution}
\end{equation}
where $S_{\rm p,pui}(\tau, t)$ describes the loss of PUIs due to neutralization on their way from point $(\tau,\mathbf{r}(\tau), w(\tau))$ to $(t,\mathbf{r}(t), w(t))$:
\begin{equation}
S_{\rm p,pui}(\tau, t) = \exp\left(- \int^t_{\tau} S_-(\hat{\tau}, \mathbf{r}(\hat{\tau}), w(\hat{\tau})) d \hat{\tau} \right).
\label{eq:sp_pui}
\end{equation}
Using Equations (\ref{eq:trajectory}) and (\ref{eq:fpui_w}) the trajectory of PUI is reconstructed backward in time from the point of phase space $(t,\mathbf{r}(t), w(t))$ to point $(t_0, \mathbf{r}(t_0), w(t_0))$, where the characteristic is close to the Sun, and it can be safely assumed that $f_{\rm pui}^{*}(t_0, \mathbf{r}(t_0), w(t_0)) = 0$. In our calculations we use the following inner boundary: $r(t_0) = R_0 = 0.1$ au.

\subsection{Kinetic moments}
In the frame of the kinetic theory, the moments of the velocity distribution function of PUIs at point $(t, \mathbf{r}) \in \mathbb{R}^4$ are the following values:
\begin{itemize}
	\item zero velocity distribution function moment -- number density:
	\begin{equation}
	n_{\rm pui}(t, \mathbf{r}) = 4 \pi \int f_{\rm pui}^*(t, \mathbf{r},w) w^2 dw;
	\label{eq:npui}
	\end{equation}

	\item second velocity distribution function moment -- kinetic temperature:
	\begin{equation}
	T_{\rm pui}(t, \mathbf{r}) = \frac{4 \pi m_{\rm p}}{3 n_{\rm pui} k_{\rm B}} \int f_{\rm pui}^*(t, \mathbf{r},w) w^4 dw, 
	\label{eq:Tpui}
	\end{equation}
	where $m_{\rm p}$ is the mass of proton, and $k_{\rm B}$ is the Boltzmann constant.
\end{itemize}
The pressure of PUIs can be calculated as $p_{\rm pui} = n_{\rm pui} k_{\rm B} T_{\rm pui}$.

\subsection{Jump condition at the TS}
%For the perpendicular or close to perpendicular parts of the TS, 
The usage of the Liouville's theorem (phase space flow conservation over the shock), the conservation of the magnetic moment (first adiabatic invariant), and assumption of the weak scattering leads to the following jump condition at the shock \citep{fahr_siewert2011, fahr_siewert2013}:
%f_{2,pui}^*(t, \mathbf{r},w) = \frac{1}{\sqrt{C}} f_{1,pui}^*(t, \mathbf{r},w/\sqrt{C}), 
%f_{\rm pui,d}^*(t, \mathbf{r},w) = \left(\frac{2s + 1}{3} \right)^{-3/2} s f_{\rm pui,u}^* \left(t, \mathbf{r}, w / \sqrt{(2s + 1) / 3} \right),
\begin{equation}
f_{\rm pui,d}^*(t, \mathbf{r}, w) = \frac{s}{C^{3/2}} f_{\rm pui,u}^* \left(t, \mathbf{r}, \frac{w}{\sqrt{C}} \right), 
\label{eq:jump-condition}
\end{equation}
where
$$
C(s, \psi) = (2 A(s, \psi) + B(s, \psi)) / 3,
$$
$$
A(s, \psi) = \sqrt{\cos^2\psi + s^2 \sin^2\psi},\: B(s, \psi) = s^2 / A^2.
$$
In Equation (\ref{eq:jump-condition}) $f_{\rm pui,u}^*$ and $f_{\rm pui,d}^*$ are the values of PUI distribution function upstream and downstream the TS, $\psi(t, \mathbf{r})$ and $s(t, \mathbf{r}) = n_{\rm d}/n_{\rm u}$ are the local upstream shock-normal angle (between the magnetic field and normal to the shock surface) and the shock compression factor that depend on the position $\mathbf{r}$ and moment $t$ of the TS crossing. From the observations by {\it Voyager 1} the compression ratio is 2.4 \citep[for TS-2 crossing, see][]{richardson2008a}, and in the global model simulations by \citet{izmod2020} it is $\approx$ 2 -- 3 over the whole solar cycle. 
%Although the TS can be considered as perpendicular at its nose and tail parts only (see Figure \ref{fig:ts}B and the discussion thereto), for the sake of simplicity we assume that Equation (\ref{eq:jump-condition}) is valid everywhere at the shock. 
From the condition (\ref{eq:jump-condition}) the downstream/upstream ratios for the moments can be obtained:
%\frac{n_{\rm pui,d}}{n_{\rm pui,u}} = s,\: \frac{T_{\rm pui,d}}{T_{\rm pui,u}} = C,\: \frac{p_{\rm pui,d}}{p_{\rm pui,u}} = s C.
\begin{equation}
	n_{\rm pui,d} / n_{\rm pui,u} = s,\: T_{\rm pui,d} / T_{\rm pui,u} = C,\: p_{\rm pui,d} / p_{\rm pui,u} = s C.
	\label{eq:mom_jump}
\end{equation}

Important to note that some PUIs can be reflected at the TS, so the reflection process leads to anisotropy (in the SW rest frame) of the velocity distribution of PUIs in some vicinity of the TS \citep{chalov2015}. In principle, the condition (\ref{eq:jump-condition}) can be modified by introducing at the TS the generation of distinct populations of PUIs (transmitted and reflected).

\subsection{Global distributions of plasma and H atoms}
To carry out calculations of the PUI distribution as described in the previous sections, the global distributions of H atoms and plasma should be known. We have performed global heliospheric simulations of SW/LISM interaction using kinetic-MHD model by \citet{izmod2020} in the stationary and time-dependent cases. Hereafter we will refer to this model as {\it IA2020}. The interstellar parameters of the models are the following: the bulk velocity and temperature are $V_{\rm LISM}$ = 26.4 km/s and $T_{\rm LISM}$ = 6530 K; the direction of $\mathbf{V}_{\rm LISM}$ is (longitude = 75.4$^\circ$, latitude = -5.2$^\circ$) in ecliptic (J2000) coordinate system; the H atom, proton, and helium ion number densities are $n_{\rm H,LISM}$ = 0.14 cm$^{-3}$, $n_{\rm p,LISM}$ = 0.04 cm$^{-3}$, and $n_{\rm He^{+},LISM}$ = 0.003 cm$^{-3}$, respectively. The SW parameters at Earth’s Orbit are the following: Mach number is 6.44 (corresponds to SW temperature $T_{\rm E}$ = 188500 K); the number density of the alpha particles $\rm He^{++}$ is 3.5\% of the proton number density. 
The heliolatitude and time variations of the SW were obtained from different experimental data ({\it OMNI 2} dataset, interplanetary scintillation data, {\it SWAN/SOHO} full-sky Lyman-$\alpha$ maps), see Appendix A in \citet[][]{izmod2020} for details. The distribution of the solar wind proton number density and velocity at 1 AU as functions of time and heliolatitude are shown in \citet[][Fig. A.1]{izmod2020}.
For the heliospheric magnetic field, the Parker spiral solution has been assumed at 1 au with magnetic field magnitude $B_{\rm E}$ = 37.5 $\mu$G at 1 au. The configuration of the interstellar magnetic field is chosen as $B_{\rm LISM}$ = 3.75 $\mu$G in magnitude and (longitude 125$^\circ$, latitude = 37$^\circ$) in direction (HGI 2000). 
The fluctuations of the heliospheric TS and the heliopause with time (in {\it Voyager 1/2} directions) are shown in \citet[][Fig. 2]{izmod2020}.
The complete description of the model can be found in \citet{izmod2015,izmod2020}.
%The only difference between {\it IA2015} and {\it IA2020} versions of the heliospheric model is in the configuration of the interstellar magnetic field: (1) $B_{\rm LISM}$ = 4.4 $\mu$G in magnitude and (longitude 163.5$^\circ$, latitude 18$^\circ$) in direction (HGI 2000 coordinate system) in {\it IA2015} model; (2) $B_{\rm LISM}$ = 3.75 $\mu$G in magnitude and (longitude 125$^\circ$, latitude = 37$^\circ$) in direction (HGI 2000) in {\it IA2020} model. 

For the charged particles, the models imply a single-fluid approach, so ``plasma" includes SW/LISM protons, pickup protons, electrons, $\alpha$ particles in SW and helium ions in LISM. In the simulations, the velocity distribution function moments of plasma (number density $n$, the bulk velocity vector $\textbf{V}$, and kinetic temperature $T$) have been calculated on specific non-regular moving grid that allows to perform exact fitting of the TS and heliopause \citep[for details, see][]{izmod2015}. Afterward, for the sake of simplicity, the kinetic moments were interpolated on the spherical grid, which is irregular by radius. For the points inside the inner boundary (1 au) we extrapolate the plasma solution with assumptions of (1) $\propto 1/r^2$ proportionality for the number density, (2) $\propto 1/r^{2(\gamma-1)}$ proportionality for the temperature (adiabatic law, $\gamma = 5/3$), and (3) linear dependence of velocity on radial distance.
Also, the obtained (using the model) time-dependent solution of the plasma distribution in the heliosphere was time-discretized and the calculations of the plasma distribution at the specific moments were performed over the entire 22-year solar cycle with a 2-month time step.

For the neutral component of H atoms, the kinetic treatment was used, and calculations were performed using the Monte-Carlo method \citep[see,][]{izmod2015}. From the global model simulations the parameters (kinetic moments) of the velocity distribution function of H atoms have been obtained everywhere in the heliosphere and interpolated on the same spherical grid as for the plasma but for the stationary case only. Additionally, to obtain the H distribution in the vicinity of the Sun more precisely the so-called two-step procedure using the local kinetic model has been used, which takes into account the solar effects \citep[for details see,][]{katushkina2010, katushkina2015}. For the time-dependent simulations, we use the model of the solar radiation pressure by \citet{kowalska2018, kowalska2020}, and for the stationary model calculations, we assume constant ratio 1.258 of the solar radiation pressure force to the solar gravitation force as it was taken in \citet{izmod2015}. The local kinetic model of H distribution uses the stationary boundary condition from the global model simulations at the sphere $r$ = 70 au. Inside the boundary sphere, the H distribution function was calculated by solving the kinetic equation with the method of characteristics, which allows taking into account non-Maxwellian properties of the H distribution in the vicinity of the Sun; the heliolatitude and time variations of the SW are considered as well. Outside the boundary sphere, the velocity distribution function of H atoms is assumed to be the sum of anisotropic (3 component) Maxwellian distributions of primary (population 4) and secondary (population 3) H atoms only. By that, we ignore the production of PUIs due to the ionization of minor populations \citep[by its relative abundance; see, ][]{malama2006,izmod2009} of H atoms that originated in the region of supersonic SW and the IHS (populations 1 and 2, respectively). The validity of such an assumption will be discussed in Section \ref{sec:results} with results.

Therefore, starting from this point, we assume that the global distributions of H atoms and plasma are known. All the following results were obtained using the distributions of plasma and H atoms calculated in the frame of the {\it IA2020} model in the stationary case unless otherwise indicated.

\subsection{Numerical calculations of the PUI distribution} 
In this section, we describe the results of calculations of PUI distribution in the heliosphere. The simulations were performed using the charged particles partitioning technique described in Section \ref{sec:partitioning}.

\begin{figure}
\includegraphics[width=\columnwidth]{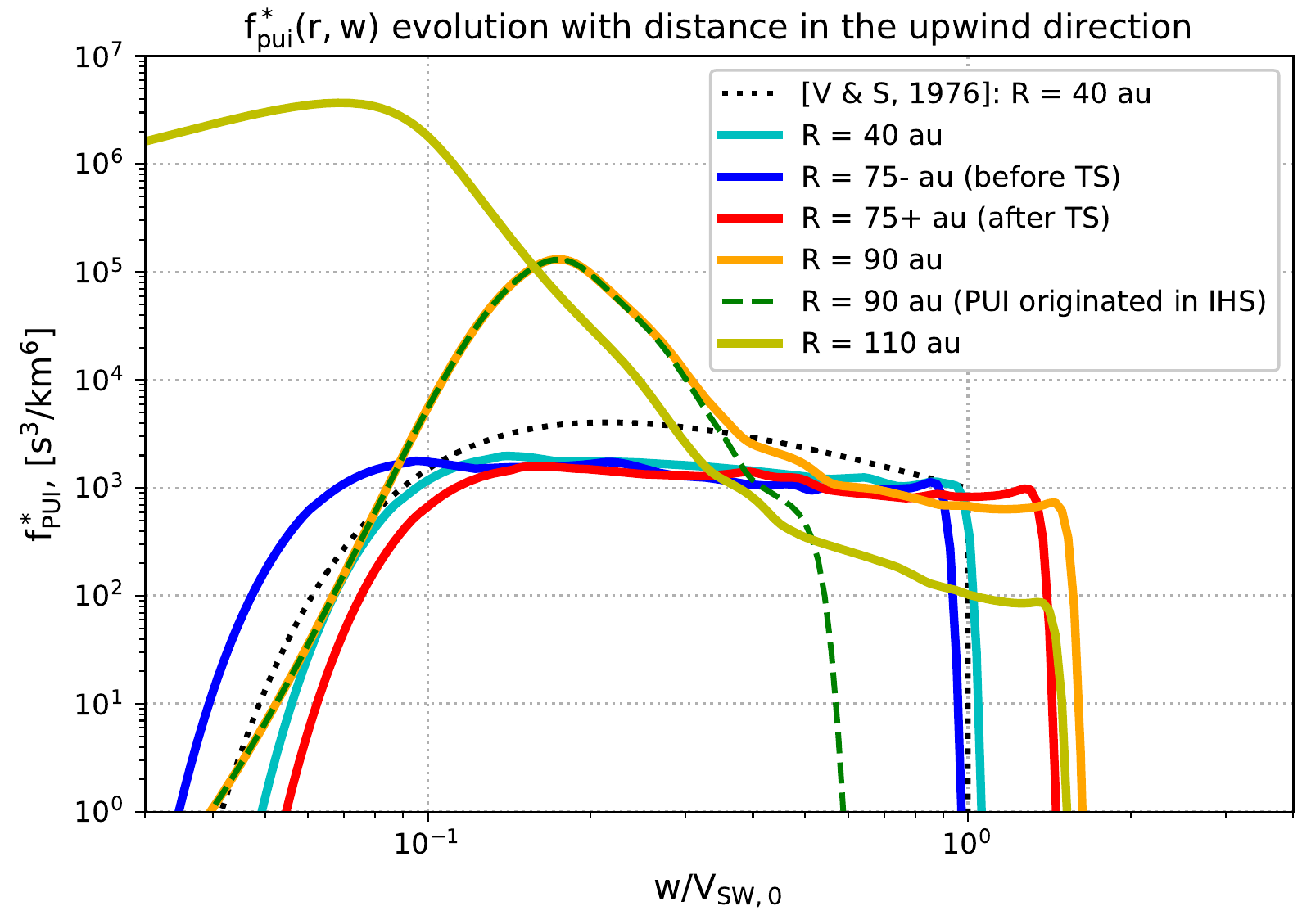}
\caption{The evolution of the PUI distribution function with distance in the upwind direction. The solid curves are the results of calculations at different heliocentric distances ($r$ = 40, 75, 90, 110 au). The blue and red solid lines are the distribution function profiles just before and after the TS, respectively (TS is located at 75 au). The green dashed curve presents the distribution function at 90 au of PUIs that originated in IHS. The black dotted line presents the analytical solution by \citet{vasyliunas1976} at $r$ = 40 au. $w$ is the velocity of PUI in the plasma reference frame, $V_{\rm sw,0}$ = 432 km s$^{-1}$.
}
\label{fig:fpui_upwind}
\end{figure}

Figure \ref{fig:fpui_upwind} presents the evolution of the PUI distribution function with distance. Numerical calculations were performed using the stationary version of the {\it IA2020} model at different heliospheric distances ($r$ = 40, 75, 90, 110 au) in the upwind (or Nose) direction (longitude $\lambda$ = 255.4$^\circ$ and latitude $\beta$ = 5.2$^\circ$ in ecliptic coordinates J2000). The velocity distribution function of PUIs is assumed to be isotropic and it depends on the velocity $w$ in the plasma reference frame.

The black dotted line of Figure \ref{fig:fpui_upwind} presents the analytical solution for PUI distribution function in the region of supersonic SW derived by \citet{vasyliunas1976}, or the so-called filled shell distribution \citep[see, also,][]{zank2010}:
$$
f^*_{\rm pui}(r, \theta, w) = \frac{3}{8 \pi} \frac{N_{\rm 0} V_{\rm 0}}{V_{\rm sw}^4} \left(\frac{V_{\rm sw}}{w}\right)^{3/2} \frac{\lambda }{r} \exp \left( -\frac{\lambda}{r} \frac{\theta}{\sin \theta} \left(\frac{V_{\rm sw}}{w}\right)^{3/2} \right),
$$
% in (Vasyliunas & Siscoe, 1976) the power in the first brakets is (3/4), not (3/2)! double-checked, it seems to be mistyped in their paper
where $w < V_{\rm sw}$, $r$ is the heliocentric distance, $\theta$ is the angle between the upwind direction and radius vector $\mathbf{r}$, $w$ is the velocity of PUI in the plasma reference frame, $N_{\rm 0}$ is the H atoms number density in the LISM, $V_{\rm 0}$ is the velocity of H atoms relative to the Sun, $V_{\rm sw}$ is the SW velocity, $\lambda = r_{\rm E}^2 \nu_{\rm ion,E} / V_{\rm 0}$ is ionization characteristic distance, and $\nu_{ion, E}$ is the ionization rate at $r_E$ = 1 au. To plot the black dotted line of Figure \ref{fig:fpui_upwind} the following set of parameters was used: $r = 40$ au, $\theta = 0$, $N_0$ = 0.14 cm$^{-3}$, $V_{\rm 0}$ = 23 km s$^{-1}$, $V_{\rm sw}$ = 432 km s$^{-1}$, $\nu_{\rm ion, E} = 6.2 \times 10^{-7}\: s^{-1}$. From Figure \ref{fig:fpui_upwind} it can be concluded that the calculations of the developed kinetic model of PUI distribution (cyan solid line) reproduces the analytical solution (black dotted line) qualitatively well. The difference in quantities can be explained by several simplifications made to derive the analytical solution, such as the assumption of a spherically expanding SW with constant velocity, the neglect of a thermal spread in velocities of SW protons and H atoms, a single neutral component assumption, the neglect of H atoms interaction with the heliospheric interface, etc.

The blue solid line of Figure \ref{fig:fpui_upwind} is the distribution function  just before (upstream) the TS (TS is located at 75 au). The difference between the maximal velocities of the cyan and blue curves, which are $\sim V_{\rm sw,0}$, can be explained by the deceleration of the SW with distance from the Sun. The red solid line of Figure \ref{fig:fpui_upwind} is the distribution function profile right after (downstream) the TS. The transition from blue to red curve represents the influence of the jump condition (\ref{eq:jump-condition}) on the PUI distribution function profile. As can be seen, after the TS crossing a PUI gets $\sqrt{C(s, \psi)} \approx \sqrt{(2s + 1) / 3}$ times higher velocity (since $\psi \approx 90^\circ$ in the upwind region), so the fast (with $w > V_{\rm sw,0}$) PUIs exist.

The green dashed curve presents the distribution function at $r$ = 90 au of PUIs that originated only in the inner heliosheath \citep[so-called ``injected" PUIs; see, e.g.,][]{zirnstein2014}. The comparison of the green and orange curves, the latter of which shows the distribution function of PUIs that originated both in the region of the supersonic SW and IHS, demonstrates that the PUIs originated in the IHS have smaller velocity in the plasma reference frame. In the IHS, the plasma flow is decelerated, and the relative velocity between the plasma and H atom, which is parental to PUI, is smaller (compared to the region of the supersonic SW).

With increasing radial distance from the Sun, the transition from the red (75 au) to orange (90 au) and yellow (110 au) curves is accompanied by a gradual decrease in the number of PUIs with $w \sim V_{\rm sw,0}$ that originated in the supersonic SW and increase in the number of slow PUIs ($w \sim 0.1 \cdot V_{\rm sw,0}$), ``injected" in the IHS. The decrease can be explained by the extinction of PUIs (due to neutralization), especially in the IHS, and it is driven by the survival probability term (\ref{eq:sp_pui}). This effect is more pronounced for PUIs with $w \sim V_{\rm sw,0}$ that originated in the supersonic SW since they travel the longer time compared to the ``injected" PUIs. The critical (or maximal) velocity of PUIs in the inner heliosheath is $w_{\rm c} \approx \sqrt{C(s, \psi)} V_{\rm sw,0}$ (in the plasma reference frame), and it is different for the red, orange, and yellow curves since the corresponding streamlines intersect slightly different regions of the TS (due to the fact that the structure of the heliosphere is perceptibly three-dimensional, so the upwind direction is not a streamline as it is in axisymmetric models), while the shock compression factor $s$ depend on the position at the TS.

\begin{figure*}
\includegraphics[width=\textwidth]{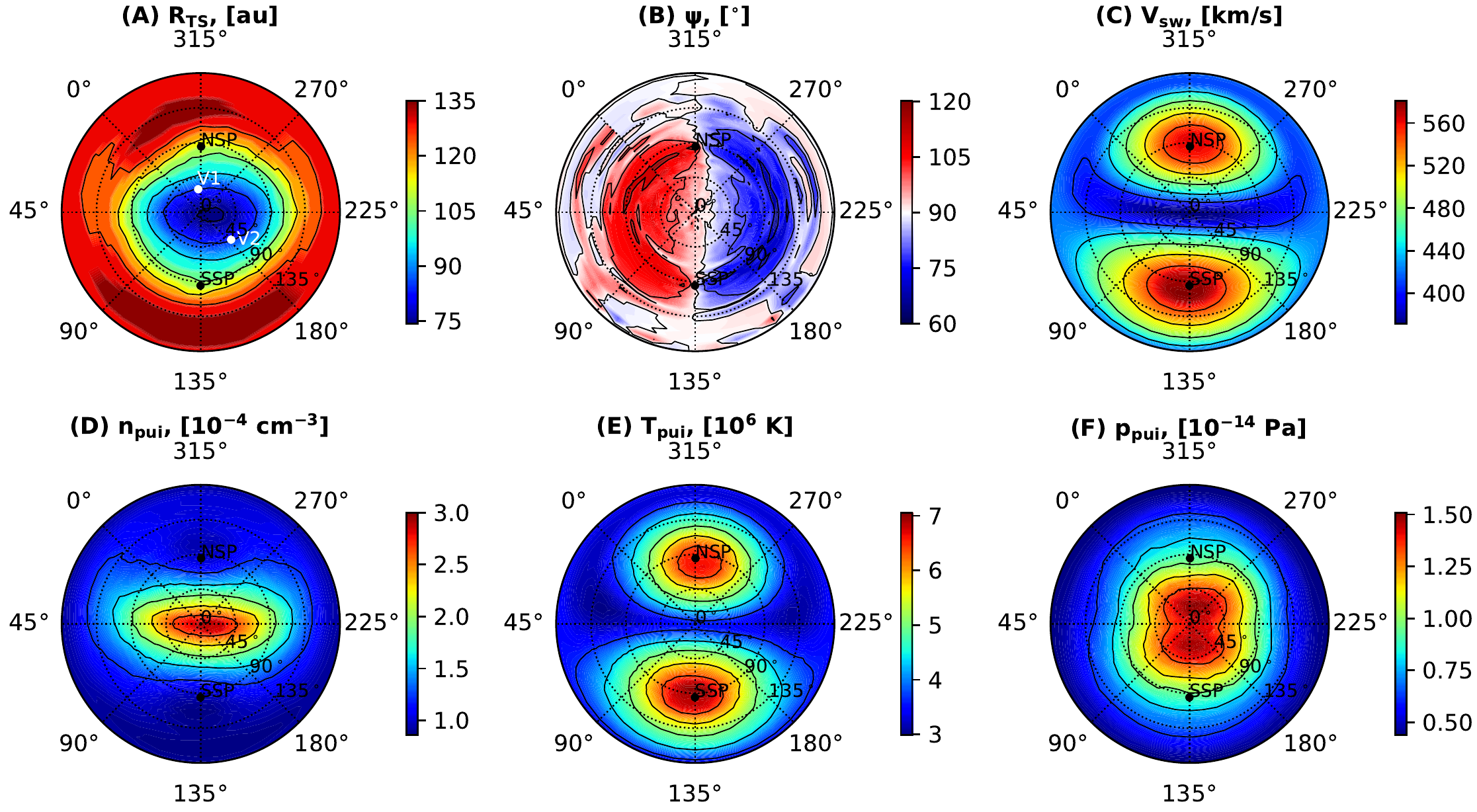}
\caption{(A) Heliocentric termination shock distance $R_{\rm TS}$. (B) The upstream shock-normal angle $\psi$. (C) The SW velocity. (D) The PUI number density. (E) The PUI kinetic temperature. (F) The PUI pressure. All the parameters are presented upstream the TS as functions of the spherical angles $\theta$, which is set in radial direction of the polar plots (in degrees), and $\varphi$ (in degrees) that is increasing in the counterclockwise direction. The definitions of $\theta$ and $\varphi$ angles are presented in the text. The values were obtained in the frame of stationary version of \citet{izmod2020} model. NSP = north solar pole, SSP = south solar pole.
\label{fig:ts}
}
\end{figure*}

Using the calculated velocity distribution function, the kinetic moments of PUIs can be obtained (see Equations \ref{eq:npui} and \ref{eq:Tpui}), which are presented in Figure \ref{fig:ts}. This figure shows the PUI moments upstream the TS: the number density (plot D), the kinetic temperature (plot E), and the pressure (plot F), as functions of the spherical angles $\theta$, which is set in radial direction of the polar plots, and $\varphi$ that is increasing in the counterclockwise direction. The angle $\theta$ is counted from the upwind direction ($\theta = 0^\circ$ direction is opposite to $\mathbf{V}_{\rm LISM}$), and the angle $\varphi$ is counted from the plane containing the $\mathbf{V}_{\rm LISM}$ and $\mathbf{B}_{\rm LISM}$ vectors (so-called BV-plane; $\varphi = 0^\circ, 180^\circ$ corresponds to BV-plane), such as projection of $\mathbf{B}_{\rm LISM}$ vector on the $(\theta = 90^\circ,\: \varphi = 0^\circ)$ direction is negative.

From Figure \ref{fig:ts}(D) we can see that the maximum of PUI number density is in the Nose region, and that PUIs are concentrated in the solar equatorial plane ($\varphi \approx 45^\circ, 225^\circ$). It can be explained by the fact, that the number density of PUIs is proportional to (a) the H atom number density $n_{\rm H}$, which has its maximum in the nose and minimum in the tail directions, and (b) the charge exchange ionization rate $\nu_{\rm ex}$, which has its minimum at solar poles \citep[see, e.g., Figure 2 from][]{katushkina2019}. The temperature of PUIs (plot E) is highly correlated with the SW velocity, which is shown in the plot C, and has its maxima in the solar pole directions since the fast SW originates from the coronal holes, which are mostly located at poles. The PUI pressure has its maximum in the upwind region (plot F).

Figure \ref{fig:ts} also presents the shape of the TS (plot A), which is elongated in the tail direction, and the upstream shock-normal angle $\psi$ (in degrees) between the magnetic field direction and outward normal to the shock (plot B). As can be seen from Figure \ref{fig:ts}(B), the TS can be considered as perpendicular (i.e., $\psi = 90^\circ$) at its nose ($\theta \approx 0^\circ$) and tail ($\theta \approx 180^\circ$) parts only, and the TS has a blunt shape in the nose region since $\psi > 90^\circ$ on the Starboard side ($\varphi \approx 45^\circ$) and $\psi < 90^\circ$ on the Port side ($\varphi \approx 225^\circ$). Therefore, it is important to note, that the jump condition at the TS (\ref{eq:jump-condition}) used in our calculations is not strictly correct everywhere at the TS, since it employs the assumption of shock perpendicularity.

\section{Charged particles partitioning} \label{sec:partitioning}

The charged particles (plasma) distribution was obtained in the frame of the {\it IA2020} heliospheric model that considers the plasma component in the context of an ideal MHD and imply a single-fluid approach. Therefore, in the calculations, the plasma represents a mixture of SW/LISM protons, pickup protons, electrons, $\alpha$ particles ($\rm He^{++}$) in SW and helium ions ($\rm He^{+}$) in LISM. For the helium ion component in the LISM and $\alpha$ particles in the SW, the continuity equations were solved separately. Then, the number density $n$ representing a mixture of protons and electrons only was obtained as $n = (\rho - m_{\rm He} n_{\rm He})/m_p$, where $\rho$ is the plasma density, $n_{\rm He}$ denotes the $\rm He^+$ number density in the interstellar medium and the $\rm He^{++}$ number density in the SW \citep[for details, see][]{izmod2015}.

To calculate the PUI distribution function, the distribution of SW protons should be known, according to Equation (\ref{eq:nu_ex_sw}). We have developed a technique to separate charged particles into components, or, to be more precise, to calculate the number density and kinetic temperature of both proton components (core SW protons and PUIs) in the IHS. This method is based on the following assumptions:		
\begin{itemize}
	\item The number density $n$ from the global simulations represents a composition of SW protons, PUIs, and electrons. To be more specific, $n = n_{\rm sw} + n_{\rm pui} + n_{\rm e} m_{\rm e} / m_{\rm p} \approx n_{\rm sw} + n_{\rm pui}$, since $m_{\rm e}/m_{\rm p} \approx 5 \times 10^{-4}$ ($m_{\rm e}$ is the mass of electron).
	\item The plasma is quasi-neutral, which leads to equation $n_{\rm e} = n_{\rm sw} + n_{\rm pui} + 2 n_{\rm He^{++}} = n + 2 n_{\rm He^{++}}$.
    \item All the populations of charged particles are co-moving, i.e. $\mathbf{V}_{\rm sw} = \mathbf{V}_{\rm pui} = \mathbf{V}_{\rm e} = \mathbf{V}_{\rm He^{++}} = \mathbf{V}$, where $\mathbf{V}$ is the plasma bulk velocity from the global calculations.
    \item For all the components the distribution functions are isotropic. The total pressure of the ionized component is equal to the sum of partial pressures, i.e. $p = p_{\rm sw} + p_{\rm pui} + p_{\rm e} + p_{\rm He^{++}}$. Therefore, the following equation can be derived
	$$
	\begin{gathered}
	(n_{\rm sw}+n_{\rm pui}+n_{\rm e}+n_{\rm He^{++}}) T = (2n + 3n_{\rm He++})T =\\
	= n_{\rm sw} T_{\rm sw} + n_{\rm pui} T_{\rm pui} + n_{\rm e} T_{\rm e} + n_{\rm He^{++}} T_{\rm He^{++}},
	\end{gathered}
	$$
	where $T$ is the plasma temperature from the global modeling.
    \item The number density of alpha particles is 3.5\% of the proton number density, i.e. $n_{\rm He^{++}} = \alpha (n_{\rm sw} + n_{\rm pui}) = \alpha n$, where $\alpha = 0.035$ \citep{izmod2015}. This assumption is not strictly correct in the whole region of the SW because near the Sun there are sources due to the ionization of the helium atoms. However, the difference rapidly decreases with distance from the Sun and becomes insignificant in the IHS, in the region of the specific interest. Also, we assume that $T_{\rm He^{++}} = T_{\rm sw}$.
    \item The temperature of electrons $T_{\rm e} = \beta T_{\rm sw}$, where $\beta = 1$ in the region of the supersonic SW as it was assumed by \citet{malama2006,chalov2013}, and $\beta = 6.7$ in the IHS \citep{chalov2019}. 
\end{itemize}

Using the assumptions above the moments of SW protons can be easily derived:
\begin{equation}
	n_{\rm sw} = n - n_{\rm pui}, T_{\rm sw} = \frac{(2 + 3\alpha) n T - n_{\rm pui} T_{\rm pui}}{(1 + \beta + (1 + 2\beta) \alpha) n - n_{\rm pui}},
	\label{eq:protons_mom}
\end{equation}

The velocity distribution function $f_{\rm sw}$ of SW protons is assumed to be isotropic Maxwellian:
$$
f_{\rm sw}(t, \mathbf{r}, \mathbf{v}) = \frac{n_{\rm sw}}{(c_{\rm sw} \sqrt{\pi})^3} \exp \left(-\frac{(\mathbf{v} - \mathbf{V}_{\rm sw})^2}{c_{\rm sw}^2}\right),
$$
where $c_{\rm sw} = \sqrt{2 k_{\rm B} T_{\rm sw} / m_{\rm p}}$ is the SW thermal velocity.

To calculate the moments of SW protons ($n_{\rm sw},\: T_{\rm sw}$) and PUIs ($n_{\rm pui},\: T_{\rm pui}$) the following iterative algorithm was used. 
\begin{enumerate}
	\item Since the PUI distribution is unknown, initially assumed that there are no PUIs ($n_{\rm pui} = 0$, $\nu_{\rm ex,pui} = 0$) and, according to Equation (\ref{eq:protons_mom}),
	$$
	n_{\rm sw} = n,\:T_{\rm sw} = \frac{2 + 3\alpha}{1 + \beta + (1 + 2\beta) \alpha} T.
	$$
	As can be seen, $T_{\rm sw} = T$ in case of $\beta = 1$ (in the supersonic SW).
	\item The simulations of the velocity distribution function of PUIs are performed as described in the previous subsections and the kinetic moments ($n_{\rm pui},\: T_{\rm pui}$) are calculated using Equations (\ref{eq:npui}) and (\ref{eq:Tpui}).
	\item Using the formulas (\ref{eq:protons_mom}) the parameters ($n_{\rm sw},\: T_{\rm sw}$) can be recalculated. 
	\item Steps ii and iii were repeated until the convergence is observed. During our numerical experiments, we have found that starting from the second iteration the number density and temperature of PUIs change insignificantly. Therefore, we have concluded that the calculations using the initial assumptions (described in Step 1) approximate the genuine distribution well.
\end{enumerate}

\begin{figure}
\includegraphics[width=\columnwidth]{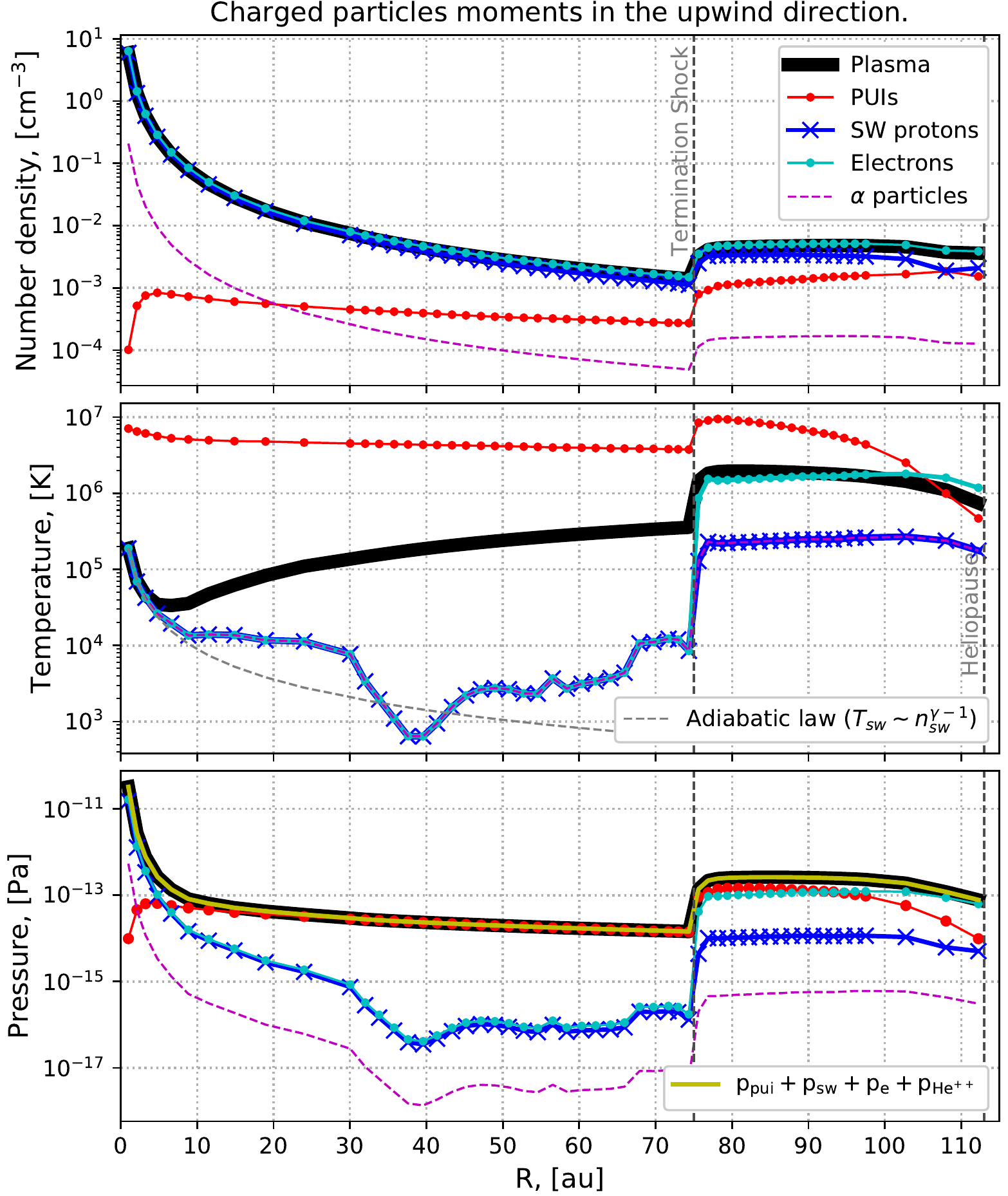}
\caption{Charged particles moments in the upwind direction. The top panel presents number densities, the middle panel -- temperatures, the bottom panel -- pressures. The plasma moments (from global simulations) have been plotted with black curves, PUIs -- red curves and dots, SW protons -- blue curves and crosses, electrons -- cyan curves and dots, $\alpha$ particles -- magenta dashed curve. For the sum of partial pressures (PUIs, SW protons, electrons, and $\alpha$ particles) the yellow curve was used. 
\label{fig:mom_iter0}
}
\end{figure}

Figure \ref{fig:mom_iter0} presents the results of the calculations of the iterative algorithm (described above) in the upwind direction. The calculations were performed for the plasma and neutral component distributions obtained in the frame of {\it IA2020} heliospheric model in the stationary case. The top panel of the figure presents the number densities, the middle panel -- the temperatures, and the bottom panel -- the pressures. The plasma distribution (from the global simulations) has been plotted with black curves, the red curves with dots were used for the PUIs, blue curves with crosses -- for the SW protons, cyan curves with dots -- for the electrons, magenta dashed curves -- for the $\alpha$ particles, and the yellow curve was used for the total pressure of PUIs, SW protons, and electrons. 

As can be seen from Figure \ref{fig:mom_iter0}, in the region of the supersonic SW the number density of PUIs remains almost constant with distance, which can be explained by the fact the production (due to ionization) and loss (due to neutralization) rates nearly compensate each other. The temperature of PUIs in this region is decreasing slowly with radial distance due to the decrease of SW velocity. In the IHS, the number density of PUIs is growing up with increasing radial distance since the production rate of the ``injected" PUIs dominates over the extinction. The temperature of PUIs in the inner heliosheath is decreasing with increasing radial distance since the fast PUIs (with $w \sim V_{\rm sw,0}$), which originated in the supersonic SW, extinct rapidly, making the distribution function more centered around small values of velocities (see Figure \ref{fig:fpui_upwind}).

The temperature of the SW protons (see blue curve in the middle panel) decreases adiabatically up to 10 AU where it becomes very low ($\sim 10^3-10^4$ K). According to \citet{malama2006}, in the region from 10 AU up to the TS the energy transferred to the SW by photoelectrons becomes non-negligible, which results in the formation of such plateau in the spatial distribution of the SW protons temperature. Downstream the TS, the temperature of the SW protons $T_{\rm sw} \approx 2 \times 10^5$ K \citep[consistent with {\it Voyager 2} observations;][]{richardson2008b}, which corresponds to SW thermal velocity $c_{\rm sw} =\sqrt{2 k_{\rm B} T_{\rm sw} / m_{\rm p}} \approx$ 58 km/s ($\approx$ 0.017 keV), and $T_{\rm sw}$ is almost constant with increasing radial distance. Since the SW thermal velocity $c_{\rm sw}$ is smaller compared to the SW velocity in the inner heliosheath ($V_{\rm sw} \approx$ 100 -- 150 km s$^{-1}$), the charge exchange of the SW protons appears to be an insignificant contributor to the measured ENA fluxes at high energies ($\gtrsim$ 1 keV). The temperature of PUIs in the inner heliosheath is by 1-2 orders of magnitude higher than the temperature of SW protons, and it is expected that the charge exchange process of PUIs is the major contributor to the ENA fluxes at these high energies. Important to note, that assumption of $\beta = 6.7$ in the IHS is essential, since in the case of $\beta = 1$ in the IHS the SW temperature downstream the TS would be $\approx$3.9 times higher in the model, making it not consistent with {\it Voyager 2} observations.

The sum of the partial pressures $p_{\rm pui}+p_{\rm sw}+p_{\rm e}+p_{\rm He^{++}}$ (the yellow curve in the bottom panel) equals the initial plasma pressure from the global simulations (black curve), which verifies the correctness and accuracy of our partitioning technique.

\section{Comparison with IBEX-Hi data} \label{sec:results}

For the comparison with our modeling results, we use the data sets of globally distributed flux observed by {\it IBEX-Hi} during the first 5 years of the mission (from 2009 to 2013) and presented by \citet{schwadron2014}, which are available on the webpage of the {\it IBEX} public Data Release 8 (\url{http://ibex.swri.edu/ibexpublicdata/Data_Release_8/}). We have calculated model full-sky maps of the ENA fluxes as described in Sections \ref{sec:fluxes}--\ref{sec:partitioning} and performed the comparison with {\it IBEX-Hi} data at the energy channels 2--6 (with the central energies $\sim$0.71, 1.1, 1.74, 2.73, and 4.29 keV, respectively). The Compton-Getting and survival probability corrections were applied for the {\it IBEX-Hi} data, so in our calculations, we did not take into account both the relative motion of the {\it IBEX} spacecraft and the ionization losses of ENAs in the region of the supersonic SW. Also, the model maps of ENA fluxes were obtained for the lines of sight when the sensor views the heliosphere in the ram direction (i.e., in the direction of spacecraft motion), as it was done for the {\it IBEX-Hi} data in \citet{schwadron2014}.

In previous works on modeling the {\it IBEX-Hi} observations, a scaling of the simulated values is applied. In this way, to perform a comparison with the {\it IBEX-Hi} data, the scaling factors 2.5 and 1.8 were used in \citet{zirnstein2017} and \citet{kornbleuth2020}, respectively. We have estimated the scaling factor for our modeling results (based on the $\chi^2$ minimization procedure; see Appendix \ref{app:scaling}) and found it to be equal to 1.002 (for the time-dependent model calculations), which is close to 1.0, so we do not scale our model results, unless otherwise indicated.

\begin{figure*}
\includegraphics[width=\textwidth]{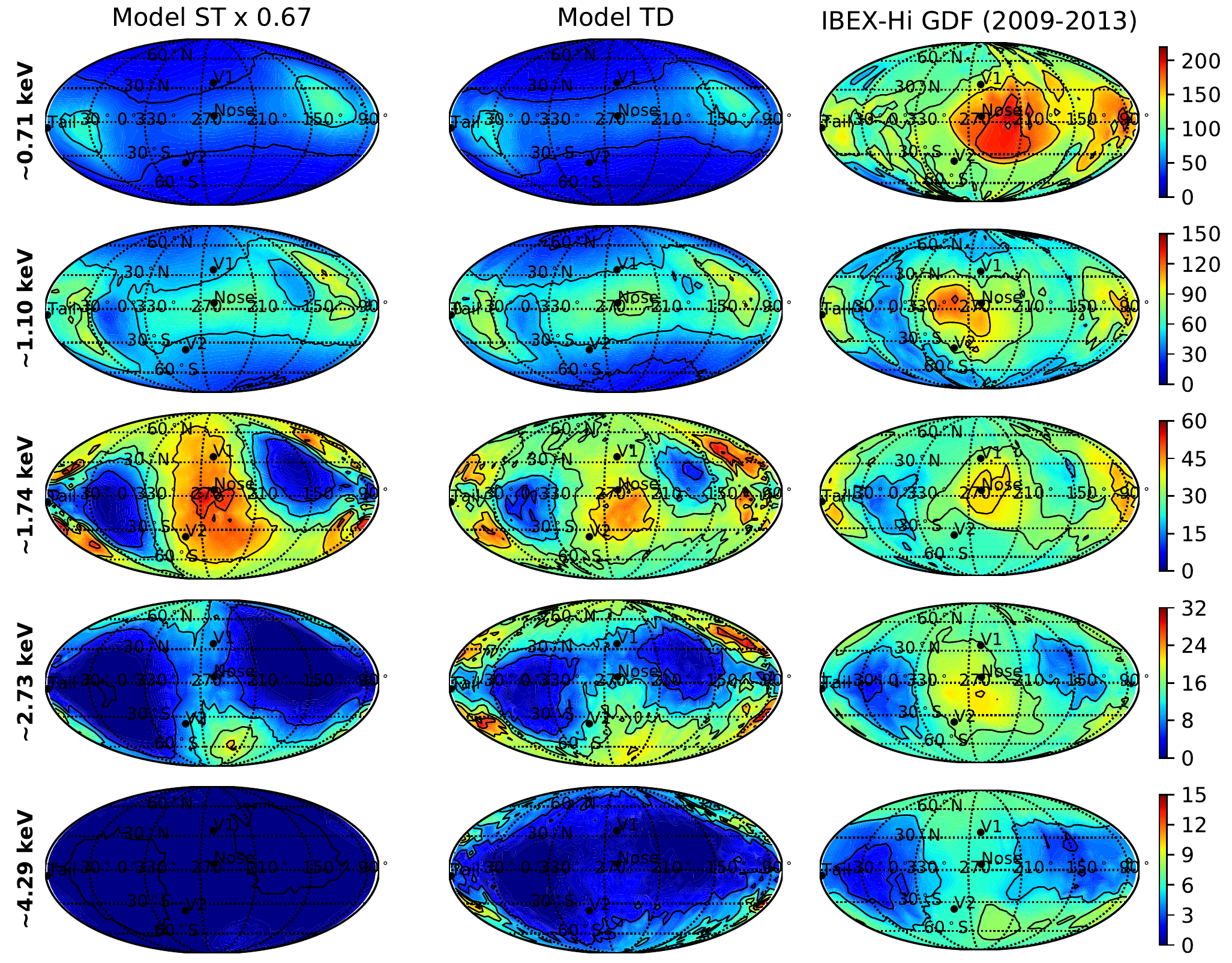}
\caption{The Mollweide skymap projections (in ecliptic J2000 coordinates) of the ENA fluxes as it was observed by {\it IBEX-Hi} at the energy channels 2--6 (by rows). The modeled ENAs originate from the PUIs only. The first column presents results of calculations using the stationary model by \citet{izmod2020}, the second column -- the model results averaged over 2009--2013 using the time-dependent model, the third column is the {\it IBEX-Hi} data collected during the same period. 
%The stationary and time-dependent model fluxes are multiplied by factors 0.66 and 0.98, respectively.
The stationary model fluxes are multiplied by factor 0.67.  
The units of fluxes are $(\rm cm^2\: sr\: s\: keV)^{-1}$. The maps are centered on the Nose ecliptic longitude $255.4^\circ$ and zero latitude. 
\label{fig:tdmaps_data_nose}
}
\end{figure*}

\begin{figure*}
\includegraphics[width=\textwidth]{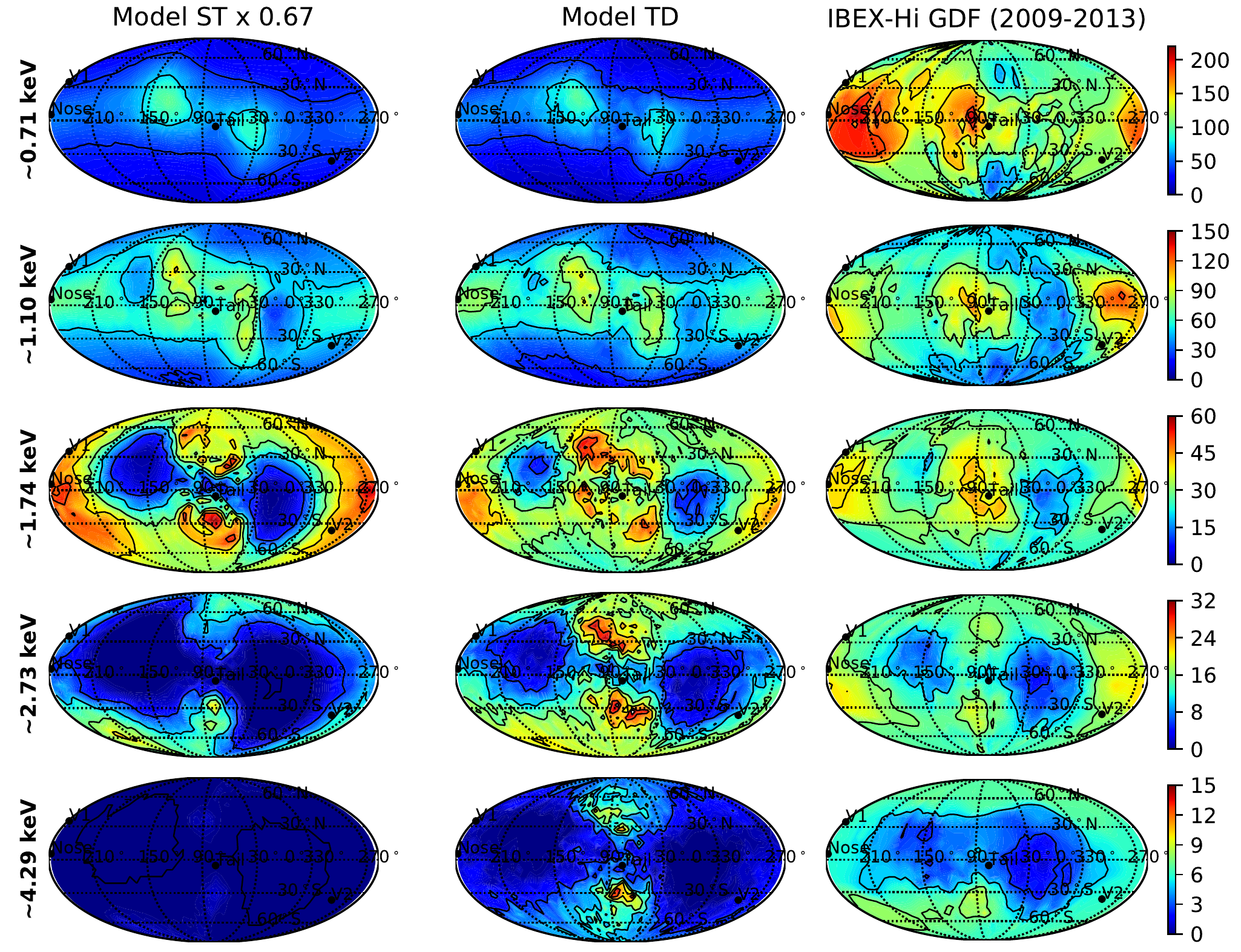}
\caption{The description is the same as for Figure \ref{fig:tdmaps_data_nose}, but the maps are centered on the Tail ecliptic longitude $75.4^\circ$.
\label{fig:tdmaps_data_tail}
}
\end{figure*}

\begin{figure*}
\includegraphics[width=\textwidth]{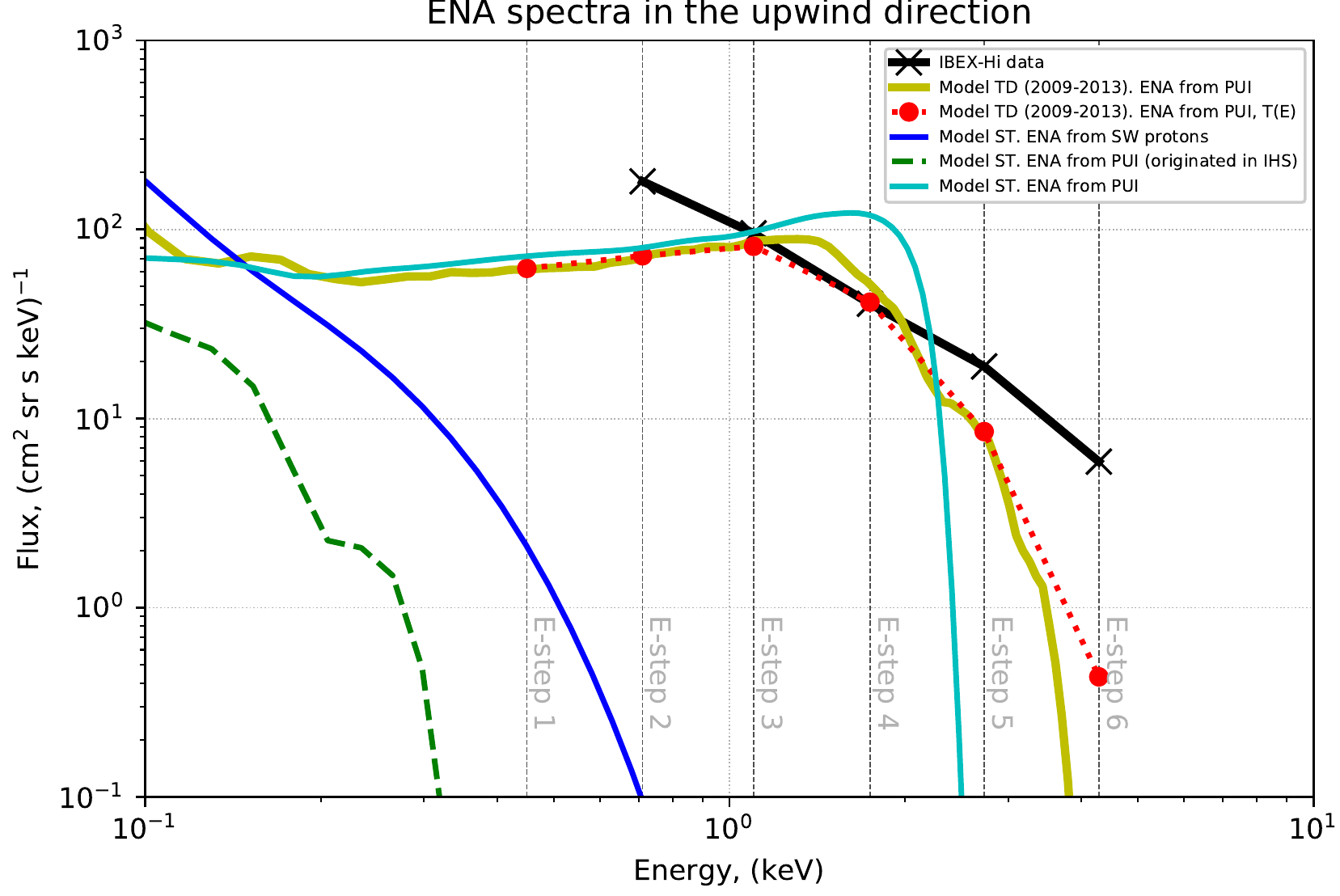}
\caption{
The ENA flux spectra as it was observed by {\it IBEX-Hi} in the upwind direction. The black solid line with crosses presents the {\it IBEX-Hi} data. The yellow solid line presents the calculated (using time-dependent model) spectrum of ENAs that originated from PUIs. The red line with dots is also the simulated fluxes of the time-dependent model, but for the specific {\it IBEX-Hi} energy steps with energy response taken into account. The blue solid line is the fluxes produced by the ENAs that originated from the SW protons (stationary results). The green dashed curve presents fluxes of ENAs that originated from PUIs that were born in IHS (stationary model). The cyan solid line is the ENA spectrum (from PUIs) calculated using the stationary model. The calculations were performed using the \citet{izmod2020} model. TD = time-dependent, ST = stationary.
\label{fig:spectra_upw}
}
\end{figure*}

\begin{figure*}
\includegraphics[width=\textwidth]{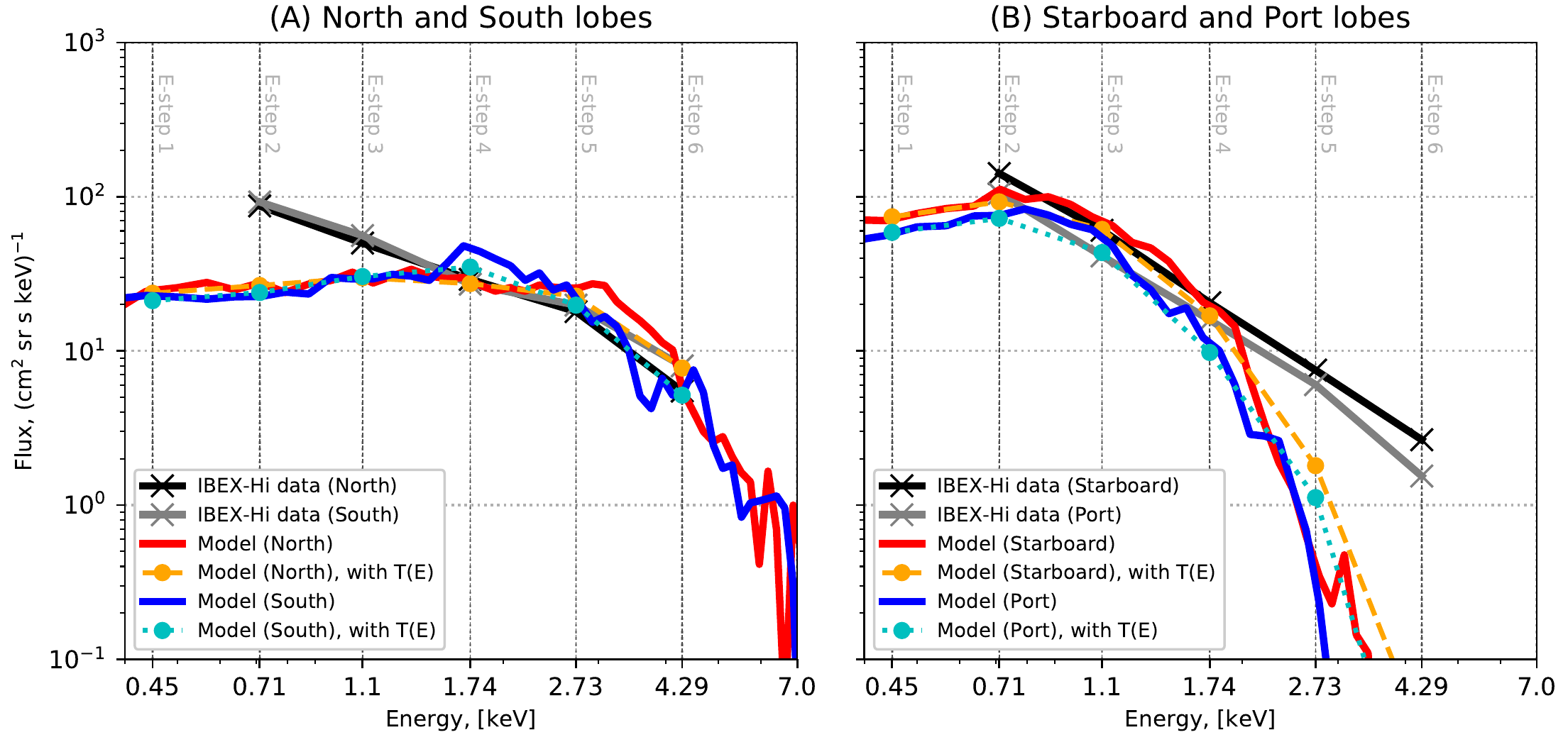}
\caption{
The ENA flux spectra as it was observed by {\it IBEX-Hi} in the directions of the North/South heliotail lobes (plot A) and the Starboard/Port lobes (plot B). The black and grey solid lines with crosses present the {\it IBEX-Hi} data, the red and blue solid curves shows the model spectra in the North/Starboard and South/Port directions, respectively. The orange and cyan dashed curves with dots are the simulated ENA fluxes for the specific {\it IBEX-Hi} energy channels (the energy response of ESAs was taken into account). The calculations were performed using time-dependent version of \citet{izmod2020} model and the results were averaged over 2009--2013. The chosen directions of the lobes are presented in Table \ref{tab:lobes}.
\label{fig:spectra_lobes}
}
\end{figure*}

The results of comparison of the {\it IBEX-Hi} data with model are presented in Figures \ref{fig:tdmaps_data_nose} -- \ref{fig:spectra_lobes}. Figures \ref{fig:tdmaps_data_nose} and \ref{fig:tdmaps_data_tail} present the full sky maps in ENA fluxes for {\it IBEX-Hi} energy channels, while Figures \ref{fig:spectra_upw} and \ref{fig:spectra_lobes} show the ENA spectra in the specific directions of the sky.

Figures \ref{fig:tdmaps_data_nose} and \ref{fig:tdmaps_data_tail} present the comparison of the {\it IBEX-Hi} data at the top five energy channels (2--6) with the ENA flux maps obtained in the frame of the {\it IA2020} heliospheric model. The first column of the figures present the results of stationary calculations, the second column -- the time-averaged (during 2009--2013) calculations using the time-dependent version of the model, the third column is the {\it IBEX-Hi} data collected during the same time period. The maps of Figure \ref{fig:tdmaps_data_nose} are centered on the Nose longitude $255.4^\circ$ and zero latitude, while the maps of Figure \ref{fig:tdmaps_data_tail} -- on the Tail longitude $75.4^\circ$ and zero latitude. 
%The stationary and time-dependent model fluxes are multiplied by corresponding best-fitting scaling factors 0.66 and 0.98, respectively.
The stationary model fluxes are multiplied by corresponding best-fitting scaling factor 0.67.  
Let us note, that the ENA fluxes are calculated for the specific {\it IBEX-Hi} energy channels with the energy response functions of electrostatic analyzers (ESAs) taken into account according to Equation (\ref{eq:flux_energy_resp}). The modeled ENAs originate from the PUI population only. We do not present in the maps the fluxes of ENAs that originated from the SW protons, since this component provide negligible fluxes in the whole {\it IBEX-Hi} energy range (2-3 orders of magnitude smaller as can be seen from Figure \ref{fig:spectra_upw}).

In Figure \ref{fig:spectra_upw} the ENA flux spectra as it was observed by {\it IBEX-Hi} in the upwind direction are presented. The black solid line with crosses shows the {\it IBEX-Hi} data. The yellow solid line presents the calculated spectrum of ENAs that originated from PUIs (using the time-dependent model). The red line with dots is also the simulated ENA fluxes of the time-dependent model, but calculated for the specific {\it IBEX-Hi} energy channels with energy response of the instrument taken into account according. The blue solid line is the model fluxes produced by the ENAs that originated from the SW protons (the stationary model was used). The green dashed curve presents the stationary model fluxes of ENAs that originated from PUIs, which were born in the IHS (``injected" PUIs). The cyan solid line is the ENA spectrum (from PUIs) calculated using the stationary version of the global heliospheric model.

Figure \ref{fig:spectra_lobes} shows the ENA flux spectra as it was observed by {\it IBEX-Hi} in the directions of the North/South heliotail regions with enhanced fluxes (plot A) and the Starboard/Port heliospheric flanks with low fluxes (plot B), or the so-called North/South and Starboard/Port lobes. In our study, these directions have taken as the center bin directions (for which the data is present) that are the closest to the lobe directions estimated by \citet[][see its Table 1]{zirnstein2016} at the fifth energy step ($\sim$2.73 keV). The chosen directions of the lobes are presented in Table \ref{tab:lobes}. The black and grey solid lines with crosses present the {\it IBEX-Hi} data, while the red and blue solid curves shows the model spectra in the North/Starboard and South/Port directions, respectively. The orange and cyan dashed curves with dots are the simulated ENA fluxes for the specific {\it IBEX-Hi} energy channels (the energy response of ESAs was taken into account).

\begin{table}
	\centering
	\caption{The directions of the heliospheric lobes in ecliptic (J2000) cordinates.}
	\label{tab:lobes}
	\begin{tabular}{lcc} % four columns, alignment for each
		\hline
		Lobe & Ecliptic & Ecliptic \\
		 	 & longitude [$^\circ$] & latitude [$^\circ$]\\
		\hline
		North & 75 & 45\\
		South & 87 & -45 \\
		Starboard & 153 & 15 \\
		Port & 9 & -15 \\
		\hline
	\end{tabular}
\end{table}

From the comparison of the simulation results with the {\it IBEX-Hi} data we can make the following conclusions:

\begin{itemize}

	\item There is a good quantitative agreement between the time-dependent model results and the observed fluxes in the middle range of energies (at energy steps 3 -- 5), especially for the regions of North/South heliotail lobes where the absolute values are well reproduced by the model even at the energy channel 6. Let us additionally note that for the simulation results, obtained in the frame of our time-dependent model, a renormalization is not needed (the scaling factor is very close to 1).
	
	\item The time-dependent version of the model explains the {\it IBEX-Hi} data better than the stationary model, especially in the Tail region (see the first and second columns of Figures \ref{fig:tdmaps_data_nose} and \ref{fig:tdmaps_data_tail}). From the comparison of the yellow and cyan curves of Figure \ref{fig:spectra_upw}, we can also see that accounting for time-dependence ``flattens" the ENA spectrum by making the fluxes lower at low ($\lesssim$2.5 keV) energies and higher at high ($\gtrsim$2.5 keV) energies, making it more consistent with the {\it IBEX-Hi} data. This can be explained by the fact that the spectrum, calculated using the time-dependent version of the model, reflect the averaged plasma properties in the IHS (during 2009 -- 2013), when the SW speed was different, which creates the spread in the ENA spectrum.
	
	\item  Both the stationary and time-dependent models are able to reproduce qualitatively the geometry of the multi-lobe structure seen in the {\it IBEX-Hi} data.
	
	A single structure of enhanced fluxes in the Tail direction is seen in the {\it IBEX-Hi} data at lower energy channels ($\sim$0.71, 1.1, 1.74 keV). At higher energies, this structure ``splits" into two parts with high fluxes (to the north and south from the solar equatorial plane). The model results qualitatively reproduce such ``splitting" behavior of the lobes. The lines of sight of the high latitude heliotail lobes with enhanced ENA fluxes intersect the regions of the IHS, where (a) the fast SW, initially emitted from the solar poles and propagated to the heliospheric tail, is collimated, and the plasma velocity and temperature are high, and (b) the heliosheath thickness is large. Figure \ref{fig:spectra_lobes}(A) shows that in the North/South lobe directions the model reproduces the data quantitatively well at all the energy steps except channel 2. In the model, the separation of the North/South heliotail lobes with enhanced fluxes is observed at $\sim$1.74 keV, while in the data the structure remains indivisible at this energy channel.

	The presence of the low flux areas from the flanks of the heliosphere (so-called Starboard and Port lobes) observed in the {\it IBEX-Hi} data is seen in the model results. The Ion and Neutral Camera ({\it INCA}) on the {\it Cassini} spacecraft \citep{krimigis2009}, which provided measurements of ENAs at high energies (5.2 -- 55 keV), has also observed these areas that were called ``basins" \citep[see, e.g.,][]{dialynas2013}. These low fluxes lobes are located in the vicinity of the solar equatorial plane, where the slow SW dominates and the thickness of the heliosheath is small. As it was suggested by \citet{mccomas2013} and studied by \citet{zirnstein2016}, the side lobes are formed by the composition of the following effects: (a) the closer the LOS to the upwind direction is, the thinner the IHS, producing lower flux; (b) in the Nose of the heliosphere, the enhancement of the ENA flux is forced by the compression and heating; (c) the emission of faster SW at high latitudes produce lobes at the flanks of the heliosphere. From Figure \ref{fig:spectra_lobes}(B) we can see that the fluxes from the Port side are systematically lower than in the Starboard region, and the model reproduces such behavior, which is due to the smaller heliosheath thickness on the Port side of the heliosphere.	
	
	As it is seen in Figure \ref{fig:spectra_lobes}, the slope of the {\it IBEX-Hi} spectrum is much smaller in the North/South lobe directions (plot A) than in the Starboard/Port lobe directions (plot B). This behavior is reproduced by the model also and can be explained by the fact that in the directions of the low latitude side lobes the lines of sight intersect those regions of the IHS, where the SW is slower and colder (with respect to the North/South lobe directions), which results in lower fluxes at high energies \citep[see, also,][]{mccomas2013}.

	\item While the model produces a comparable with the {\it IBEX-Hi} data fluxes at energy channels $\sim$1.1 keV and $\sim$1.74 keV, a deficit of fluxes at higher energy channels 5 and 6 is observed, especially from the Nose region. From Figure \ref{fig:spectra_upw} we see that the {\it IBEX-Hi} data values (black crosses) at the energy channels 5 and 6 are $\sim$2 and $\sim$10 times, respectively, higher than the model fluxes (red points). The possible explanation for this discrepancy is a lack of ENAs, which originated from 
the energetic PUI population produced by the processes of shock-drift acceleration or reflection from the cross-shock potential.
%the PUIs that initially reflected from the TS, gained some amount of energy afterward, and then transmitted through the TS (so-called reflected PUI population). 
%In the Nose direction, the TS is almost perpendicular, so the reflection process of PUIs is effective the most. 
The presence of such energetic PUIs is not taken into account in our modeling.
	
	\item The model produces smaller fluxes (compared to the {\it IBEX-Hi} data) from both the Nose and Tail regions of the sky at energy channel 2 (central energy $\sim$0.71 keV). As can be seen in Figure \ref{fig:spectra_upw}, neither the ENAs that originated from the SW protons (blue solid curve) nor the ENAs that were created through the charge exchange of ``injected" PUIs (green dashed line), which originated in the inner heliosheath, can explain the discrepancy between the model calculations and the {\it IBEX-Hi} data at this energy channel since the atoms of these populations have quite low energy ($\lesssim$0.1 keV). 
	
	Important to note, that in our simulations the origin of the PUIs in result of charge exchange of protons with H atoms, which originated in the supersonic SW or IHS (so-called populations 1 and 2 of H atoms, respectively), is not taken into account. In principle, the atoms of population 1 have velocity ${\sim}V_{\rm sw}$ \citep[for details see,][]{izmod2009} and potentially they could be a source of PUIs that will be parental to ENAs with energies $\lesssim$1 keV. The atoms of population 2 could be a seed population for high energy PUIs, which, in turn, will be parental to ENAs with few keV energies. Nevertheless, as was shown by \citet[][see Figure 6]{malama2006}, in the {\it IBEX-Hi} energy range these scenarios produce ENA fluxes, which are 1-2 orders smaller than the fluxes from the populations considered in our modeling, so they can be safely neglected. According to the results of \citet{malama2006} there are no other H atom populations that can explain the lack of ENAs in the modeling at {\it IBEX-Hi} energy channel 2 ($\sim$0.71 keV).
	
	The possible explanation for the lack of fluxes at low energies ($\sim$0.7-1 keV) can be the velocity diffusion, which is assumed negligible in our modeling. The acceleration effect driven by the velocity diffusion of PUIs in the inner heliosheath may produce higher fluxes at $\sim$1 keV energies \citep[see, e.g.,][]{kallenbach2005, fahr2011, fahr2016}. 

\end{itemize}

\section{The effect of additional energetic PUI population on ENA fluxes} \label{sec:tail}

The main limitation of the described above model is the absence of processes that produce energetic ``tails'' in PUI distribution, such as shock-drift acceleration or reflection from the cross-shock potential (``shock-surfing'' mechanism). To see how the inclusion of additional energetic PUIs affects the modeled ENA fluxes, we consider a ``toy model" with a power-law ``tail" in PUI distribution downstream the TS.

To simulate the additional energetic PUI population, the following approach will be employed. Using the method described in Section \ref{sec:model}, the PUI velocity distribution function can be calculated everywhere downstream the TS. This distribution is the filled shell $f_{\rm sh}^{*}(t, \mathbf{r}, w)$ with critical velocity $w_{\rm c} \approx \sqrt{C(s, \psi)} \cdot |\mathbf{V}_{\rm sw} - \mathbf{V}_{\rm H}|$, i.e. $f_{\rm sh}^{*}(t_{\rm TS}, \mathbf{r}_{\rm TS}, w) = 0$ for $w > w_{\rm c}$, where $s$ is the shock compression factor, $\mathbf{V}_{\rm sw}$ is the SW velocity vector, $\mathbf{V}_{\rm H}$ is H atom bulk velocity, $t_{\rm TS}$ and $\mathbf{r}_{\rm TS} = \mathbf{r}(t_{\rm TS})$ are the moment and position of a PUI crossing the TS. For velocities higher than $w_{\rm c}$ we assume the power law distribution $f_{\rm tail}^{*}(t_{\rm TS}, \mathbf{r}_{\rm TS}, w) \sim w^{-\eta}$ with index $\eta$ that defines the slope of the ``tail". We also introduce the additional parameter $\xi$ that is the density fraction of the PUIs of the ``tail" distribution. Therefore, the PUI velocity distribution function right after the TS is assumed as
\begin{equation}
f_{\rm pui, d}^{*} =(1 - \xi) f_{\rm sh}^{*} + \xi f_{\rm tail}^{*},
\label{eq:bound_cond}
\end{equation}
where
\begin{equation}
\begin{split}
f_{\rm tail}^{*}(t_{\rm TS}, \mathbf{r}_{\rm TS}, w) 
&= n_{\rm pui,d}(t_{\rm TS}, \mathbf{r}_{\rm TS}) \frac{w^{-\eta}}{4 \pi \int_{w_{\rm c}}^{+\infty} w^{-\eta} w^2 dw} \\
&= n_{\rm pui,d}(t_{\rm TS}, \mathbf{r}_{\rm TS}) \frac{\eta - 3}{4 \pi w_{\rm c}^3} \left(\frac{w}{w_{\rm c}}\right)^{-\eta},\: w \geq w_{\rm c},
\end{split}
\label{eq:ftail}
\end{equation}
and $f_{\rm tail}(t_{\rm TS}, \mathbf{r}_{\rm TS}, w) = 0$ for $w < w_{\rm c}$, $n_{\rm pui,d}$ is the PUI number density downstream the TS. Equation (\ref{eq:ftail}) implies that $\eta > 3$, otherwise the number density of the ``tail" PUIs will be infinite.

In the IHS, the solution of the kinetic equation (\ref{eq:solution}) with the boundary condition downstream the TS (\ref{eq:bound_cond}) is employed. The first and second terms of this condition are associated with transmitted and additional energetic populations of PUIs downstream the TS, respectively. Let us note that such partitioning of PUIs (into sub-populations) implies the conservation of the total number density, while the pressure balance is not satisfied. By introducing the ``tail" in the velocity distribution, some amount of energy is added to the system. This additional energy can be ``pumped" by the interaction of PUIs with the TS or produced by fluctuations of the heliospheric magnetic field.

\begin{figure*}
\includegraphics[width=\textwidth]{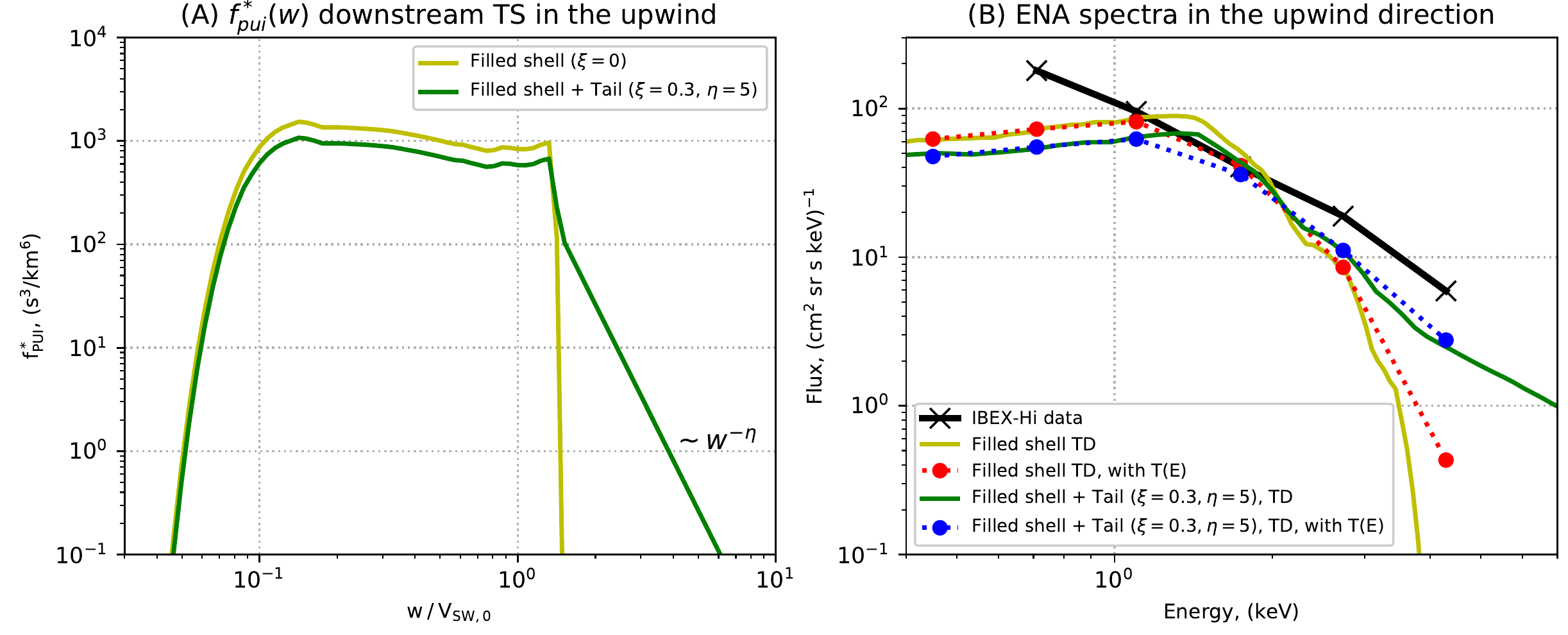}
\caption{
(A) The velocity distribution function of PUIs downstream the TS and (B) the ENA flux spectra as it was observed by {\it IBEX-Hi} in the upwind direction, calculated with different assumptions on PUI distribution right after the TS: the yellow curve -- filled shell distribution; the green curve -- the sum of the filled shell distribution and power law ``tail" with parameters $\xi = 0.3$ and $\eta = 5$ (the definition of the parameters can be found in the text). The red and blue dots are the simulated ENA fluxes for the specific {\it IBEX-Hi} energy channels. The black solid line with crosses presents the {\it IBEX-Hi} data. The simulations of the PUI distribution functions (plot A) were performed in the stationary case, and the spectra (plot B) are calculated using the time-dependent model by \citet{izmod2020}. $V_{\rm sw,0}$ = 432 km s$^{-1}$. TD = time-dependent.
}
\label{fig:fpui_spectra_tail}
\end{figure*}

\begin{figure*}
\includegraphics[width=0.49\textwidth]{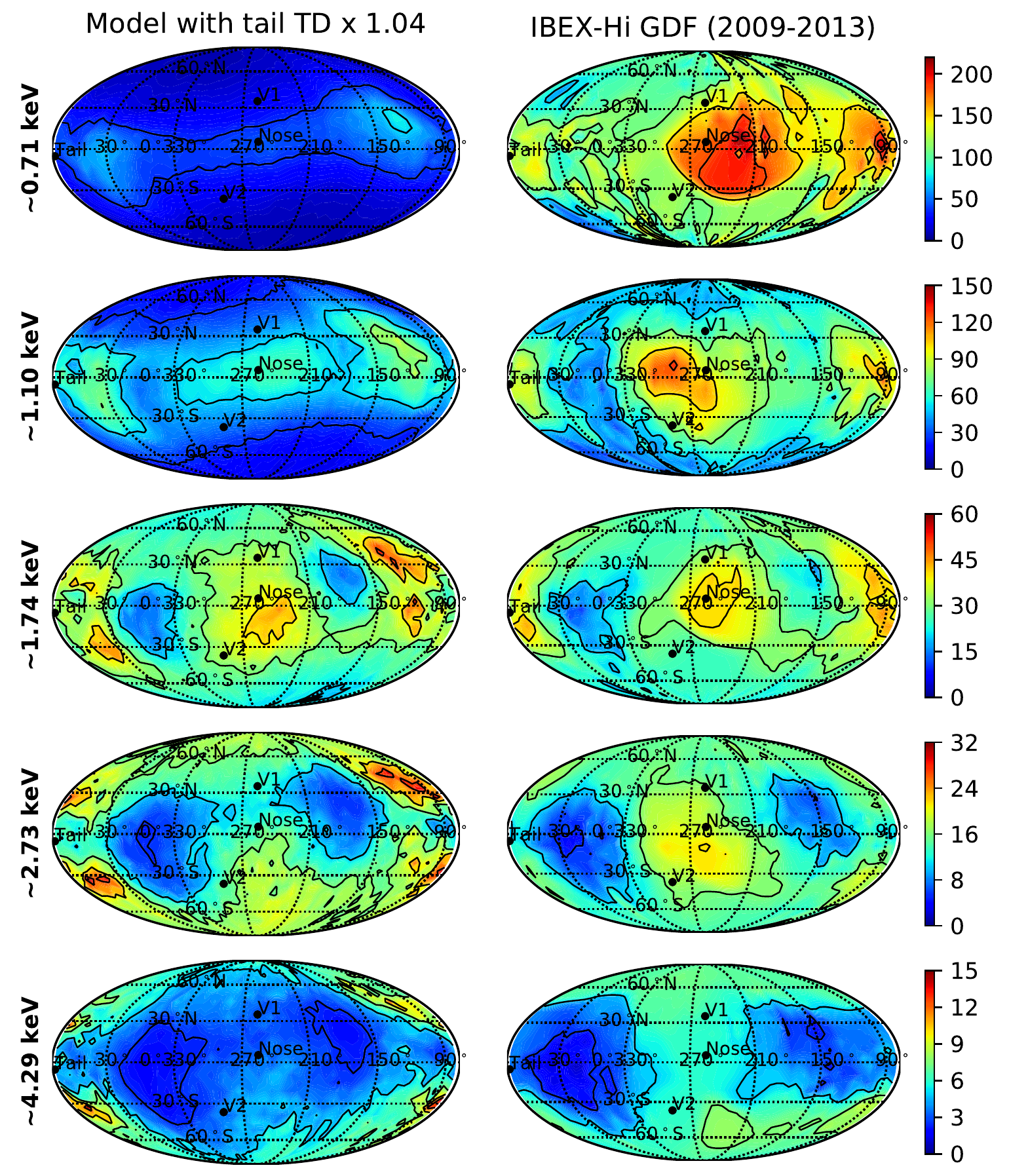}
\includegraphics[width=0.49\textwidth]{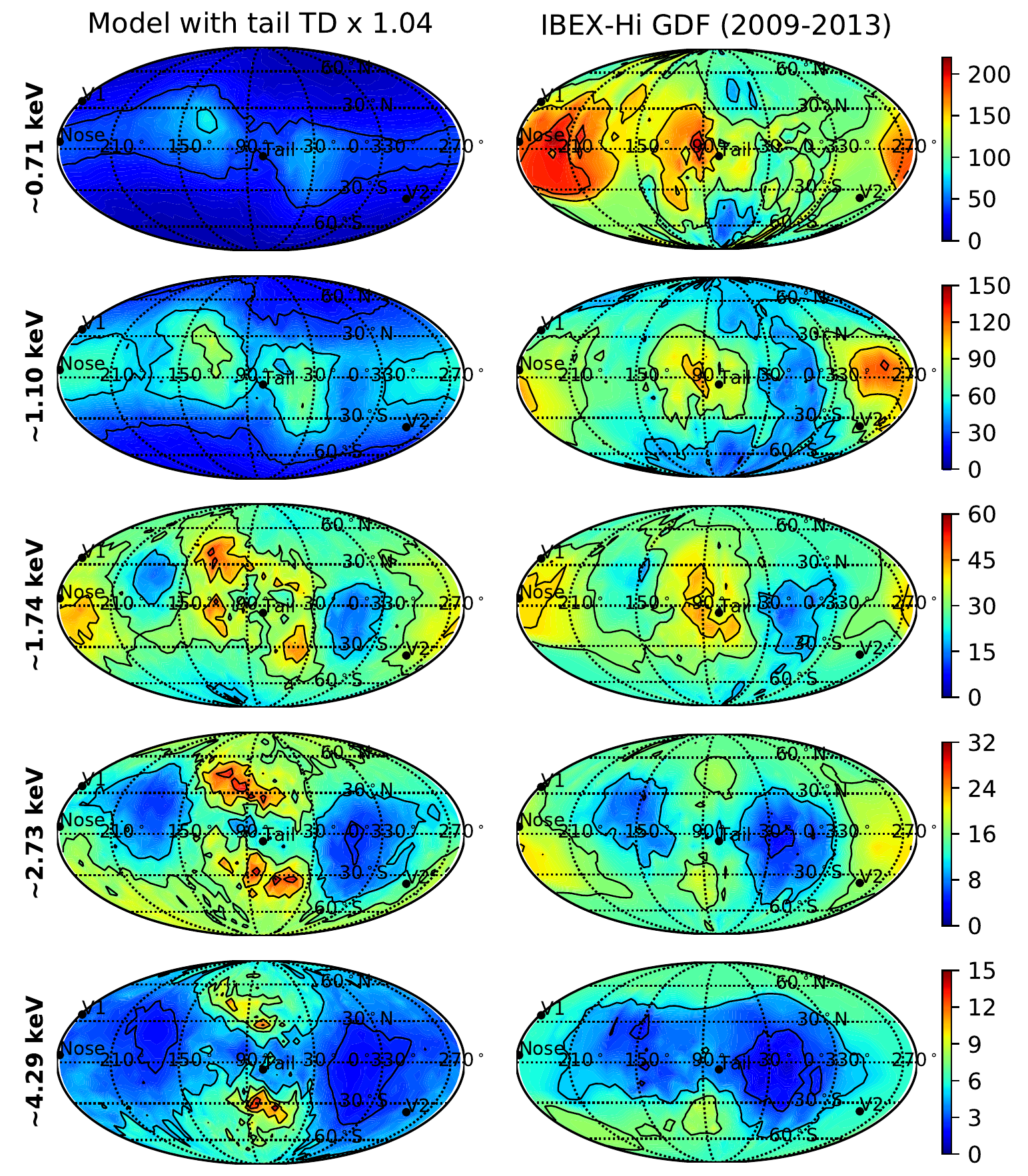}
\caption{
The Mollweide skymap projections (in ecliptic J2000 coordinates) of the ENA fluxes as it was observed by {\it IBEX-Hi} at the energy channels 2--6 (by rows). The modeled ENAs originate from the PUIs only. The first and third columns present the results of calculations using the time-dependent model by \citet{izmod2020} with the ``tail" in the PUI distribution downstream the TS ($\xi = 0.3,\: \eta = 5$), the second and fourth columns are the {\it IBEX-Hi} data collected during 2009 -- 2013. The model fluxes were multiplied by the best-fitting scaling factor $\hat k$ = 1.04. The maps in the first and second columns are centered on the Nose longitude $255.4^\circ$, and the maps in third and fourth columns -- on the Tail longitude $75.4^\circ$.
}
\label{fig:powertail}
\end{figure*}

In principle, the parameters $\xi$ and $\eta$, which were artificially introduced in the approach described above, depend on the local TS properties, such as shock-normal angle $\psi$ and shock compression factor $s$, but for the sake of simplicity, we assume it to be constant in our study. In order to demonstrate the qualitative effect of the additional energetic PUIs on the ENA fluxes, we have performed calculations for the specific pair of parameters -- $\xi = 0.3$ and $\eta = 5$ \citep{fisk2007}. 

Figure \ref{fig:fpui_spectra_tail} shows the velocity distribution function of PUIs downstream the TS (plot A) and the ENA flux spectra as it was observed by {\it IBEX-Hi} in the upwind direction (plot B), calculated with different assumptions on PUI distribution right after the TS. The yellow curves present the results of calculations using the filled shell distribution downstream the TS, and the green curve -- the sum of the filled shell distribution and power-law ``tail" (with parameters $\xi = 0.3$ and $\eta = 5$). The red and blue dotted lines are the simulated ENA fluxes for the specific {\it IBEX-Hi} energy channels with the energy response of ESAs taken into account. The black solid line with crosses presents the data. Figure \ref{fig:powertail} shows the comparison of the {\it IBEX-Hi} data at the top five energy channels (2--6) with the simulated full-sky ENA maps in the frame of the time-dependent version of {\it IA2020} heliospheric model with the ``tail" in the PUI distribution downstream the TS.

As can be seen from Figures \ref{fig:fpui_spectra_tail} and \ref{fig:powertail}, the additional population of energetic PUIs (simulated in our approach using the power-law ``tail" in the PUI distribution) produce higher fluxes at the top energy channels, which makes the model flux maps qualitatively and quantitatively more consistent with the {\it IBEX-Hi} data. As can be concluded, the GDF is extremely sensitive to the form of the velocity distribution function of PUIs in the inner heliosheath, and the accounting for the existence of additional energetic population of PUIs is essential to explain the data. Therefore, a detailed parametric study of the this population using the {\it IBEX-Hi} data needs to be performed, which is beyond the scope of this paper. It is planned to be done in the future and will be published elsewhere.

\section{Conclusions} \label{sec:conclusions}

In this work, we have calculated the ENA fluxes at Earth’s orbit and performed a detailed quantitative comparison with the {\it IBEX-Hi} data. The main conclusions of these studies can be summarized as follows.

\begin{enumerate}
	
	\item In the model described in this paper, the PUI population is considered kinetically. Using the developed model, we were able to calculate the full-sky ENA flux maps and reproduce the geometry of the multi-lobe structure seen in the {\it IBEX-Hi} data.  There is a good quantitative agreement between the time-dependent model results and the observed fluxes in the middle range of energies (at energy steps 3 -- 5), especially for the regions of North/South heliotail lobes, where the absolute values are well reproduced by the model even at the energy channel 6. For the time-dependent model results a scaling of the fluxes is not needed.
	
	\item Despite a relatively good agreement, there are few quantitative differences between our model calculations and the {\it IBEX-Hi} data: 
	(a) a deficit of fluxes at highest energy channels 5 and 6 is observed, especially from the Nose region;
	(b) the model produces smaller fluxes (compared to the {\it IBEX-Hi} data) from both the Nose and Tail regions of the sky at energy channel 2 (central energy $\sim$0.71 keV); 
	(c) the ``split" of the North/South heliotail regions with enhanced fluxes is observed in the model at lower energies ($\sim$1.5 keV) than in the data ($\sim$2 keV). 
	These distinctions can be the result of several assumptions and simplifications made in the modeling, such as the isotropic form of the velocity distribution function of PUIs everywhere in the heliosphere, the neglect of the velocity diffusion, and the weak pitch-angle scattering at the TS.

	\item The ENA fluxes from the inner heliosheath are extremely sensitive to the form of the PUI velocity distribution function. The accounting for the existence of additional energetic population of PUIs is essential to explain the data.	
	
\end{enumerate}

Thereby, the goal of future investigations is to take into account the additional population of energetic PUIs and understand the physical reasons for the lack of ENA fluxes at energy channel 2 ($\sim$0.7 keV). For these purposes, we plan to consider the realistic dynamics of PUIs near the TS. Some portion of PUIs can experience reflections at the shock front due to abrupt change of the magnetic field and gain energy from their drift motion along the TS in the direction of the induced electric field. The PUIs can also experience the pitch-angle scattering upstream and downstream of the TS, which provides a way for transmitted particles to return to the shock, so the multiple reflections can occur. The process of reflection leads to anisotropy (in the SW rest frame) of the velocity distribution of PUIs near the TS. Thus, in this vicinity (upstream and downstream of the TS), the transport equation for anisotropic velocity distribution function of PUIs should be solved \citep[see, e.g.,][]{chalov2015}.

\section*{Acknowledgements}
The authors would like to acknowledge Dr. Nathan Schwadron for providing information on IBEX GDF uncertainties.
The work of I.I. Baliukin and V.V. Izmodenov was supported by the Foundation for the Advancement of Theoretical Physics and Mathematics ``BASIS" 18-1-1-22-1.
The numerical single-fluid MHD code of the SW/LISM global interaction used in this paper was developed by D.B. Alexashov in the frame of the Russian Science Foundation grant 19-12-00383.
The authors would like to thank for the discussions that appeared during online meetings in the frame of NASA 18-DRIVE18\_2-0029, Our Heliospheric Shield, 80NSSC20K0603 project. The authors are no financial support collaborators of the DRIVE project.

%The authors would like to thank the support of NASA grant 18-DRIVE18\_2-0029, Our Heliospheric Shield, 80NSSC20K0603.

\section*{Data availability}
The data underlying this article will be shared on reasonable request to the corresponding author.

%%%%%%%%%%%%%%%%%%%%%%%%%%%%%%%%%%%%%%%%%%%%%%%%%%



\bibliographystyle{mnras}

\begin{thebibliography}{99}
%\bibitem[Chalov \& Fahr (1996)]{chalov1996} Chalov \& Fahr 1996
\bibitem[Chalov \& Fahr (1997)]{chalov1997} Chalov, S. V., \& Fahr, H. J. 1997, \aap, 326, 860-869
\bibitem[Chalov et al. (2003)]{chalov2003} Chalov, S. V., Fahr, H. J., \& Izmodenov, V. V. 2003, J. Geophys. Res., 108, 1266
%\bibitem[Chalov (2005)]{chalov2005} Chalov 2005
\bibitem[Chalov (2010)]{chalov2010} Chalov, S. V., Alexashov, D. B., McComas, D., et al. 2010, ApJL, 716, L99
\bibitem[Chalov \& Fahr (2013)]{chalov2013} Chalov S.V., \& Fahr H.J., 2013, MNRAS, 433, L40
\bibitem[Chalov et al. (2015)]{chalov2015} Chalov S.V., Malama, Y. G., Alexashov, D.B., Izmodenov V.V., 2015, MNRAS, 455, 431-437
\bibitem[Chalov (2019)]{chalov2019} Chalov, S.V. 2019, MNRAS, 485, 4, 5207-5209
\bibitem[Dialynas (2013)]{dialynas2013} Dialynas, K., Krimigis, S. M., Mitchell, D. G., Roelof, E. C., \& Decker, R. B. 2013, \apj, 778, 40
%\bibitem[Fahr \& Lay (2000)]{fahr2000} Fahr \& Lay 2000
\bibitem[Fahr \& Chalov (2008)]{fahr2008} Fahr, H.J., \& Chalov, S.V. 2008, \aap, 490, L35-L38
\bibitem[Fahr \& Fichtner (2011)]{fahr2011} Fahr, H. J., \& Fichtner, H. 2011, \aap, 533, A92+
\bibitem[Fahr \& Siewert (2011)]{fahr_siewert2011} Fahr, H. J., \& Siewert, M. 2011, \aap, 527, A125
\bibitem[Fahr \& Siewert (2013)]{fahr_siewert2013} Fahr, H. J., \& Siewert, M. 2013, \aap, 552, A41
\bibitem[Fahr et al. (2016)]{fahr2016} Fahr, H. J., Sylla, A., Fichtner, H., and Scherer, K. 2016, J. Geophys. Res. Space Physics, 121, 8203-8214
\bibitem[Fisk \& Gloeckler (2007)]{fisk2007} Fisk, L. A. \& Gloeckler, G. 2007, Proceedings of the National Academy of Sciences, 104, 14, 5749-5754
\bibitem[Funsten et al. (2009)]{funsten2009} Funsten, H. O., Allegrini, F., Bochsler, P. et al., 2009, SSRv, 146, 75--103
\bibitem[Gloeckler et al. (1994)]{gloeckler1994} Gloeckler, G., Geiss, J., Roelof, E.C. et al. 1994, JGR: Space Physics, 99, A9, 17,637-17,643
\bibitem[Heerikhuisen et al. (2008)]{heerikhuisen2008} Heerikhuisen, J., Pogorelov, N. V., Florinski, V., Zank, G. P., \& le Roux, J. A. 2008, \apj, 682, 679
\bibitem[Heerikhuisen et al. (2010)]{heerikhuisen2010} Heerikhuisen, J.,  Pogorelov, N. V., Zank, G. P. et al. 2010, \apj, 708, L126
\bibitem[Isenberg (1987)]{isenberg1987} Isenberg, P. A. 1987, JGR, 92, 1067
\bibitem[Izmodenov (2001)]{izmod2001} Izmodenov, V.~V., The Outer Heliosphere: The Next Frontiers, Edited by K. Scherer, Horst Fichtner, Hans J{\"o}rg Fahr, and Eckart Marsch COSPAR Colloquiua Series, 11. Amsterdam: Pergamon Press, 2001, 23
\bibitem[Izmodenov et al. (2009)]{izmod2009} Izmodenov, V.V., Malama , Y.G., Ruderman, M.S., et al., Kinetic-Gasdynamic Modeling of the Heliospheric Interface, SSRv, 2009, 146, 329-351
\bibitem[Izmodenov \& Alexashov (2015)]{izmod2015} Izmodenov V.V., \& Alexashov D.B., 2015, \apjs, 220, 32
\bibitem[Izmodenov \& Alexashov (2020)]{izmod2020} Izmodenov V.V., \& Alexashov D.B., 2020, \aap, 633, L12
\bibitem[Kallenbach et al. (2005)]{kallenbach2005} Kallenbach, R., Hilchenbach, M., Chalov, S. V., Le Roux, J. A., and Bamert, K. 2005, \aap, 439, 1-22
\bibitem[Katushkina \& Izmodenov (2010)]{katushkina2010} Katushkina, O. A., \& Izmodenov, V. V. 2010, AstL, 36, 297
\bibitem[Katushkina et al. (2015)]{katushkina2015} Katushkina, O. A., Izmodenov, V. V., Alexashov, D. B., Schwadron, N. A., and McComas, D. J. 2015, \apjs, 220, 33
\bibitem[Katushkina et al. (2019)]{katushkina2019}  Katushkina, O. A., Izmodenov V. V., Koutroumpa, D., Quemerais, E., Jian, L. K., 2019, Solar Phys., 294, 17
\bibitem[Kornbleuth et al. (2018)]{kornbleuth2018} Kornbleuth M., Opher M., Michael A.T., Drake J.F. 2018, \apj, 865, 84
\bibitem[Kornbleuth et al. (2020)]{kornbleuth2020} Kornbleuth M., Opher M., Michael A.T., et al. 2020, ApJL, 895, L26
\bibitem[Kowalska-Leszczynska et al. (2018)]{kowalska2018} Kowalska-Leszczynska, I., Bzowski, M., Sokol, J. M., Kubiak, M. A. 2018, \apj, 852, 2, 115
\bibitem[Kowalska-Leszczynska et al. (2020)]{kowalska2020} Kowalska-Leszczynska, I., Bzowski, M., Kubiak, M. A., Sokol, J. M. 2020, \apjs, 247, 2, 62
\bibitem[Krimigis et al. (2009)]{krimigis2009} Krimigis, S. M., Mitchell, D. G., Roelof, E. C., Hsieh, K. C., \& McComas, D. J. 2009, Science, 326, 971
\bibitem[Lindsay \& Stebbings (2005)]{lindsay2005} Lindsay, B. G., \& Stebbings, R. F. 2005, JGRA, 110, A12213
\bibitem[McComas et al. (2009)]{mccomas2009} McComas, D. J., Allegrini, F., Bochsler, P., et al. 2009, Science, 326, 959
\bibitem[McComas et al. (2012)]{mccomas2012} McComas, D.~J., Alexashov, D.~B., Bzowski, M., et al.\ 2012, Science, 336, 6086, 1291
\bibitem[McComas et al. (2013)]{mccomas2013} McComas, D.~J., Dayeh, M. A., Funsten,  H. O. et al. 2013, \apj, 771, 77
%\bibitem[McComas et al. (2014)]{mccomas2014} McComas, D. J., Allegrini, F., Bzowski M., et al. 2014, ApJS, 213, 20
\bibitem[McComas et al.(2015)]{mccomas2015} McComas, D.~J., Bzowski, M., Frisch, P., et al.\ 2015, \apj, 801, 28
\bibitem[McComas et al.(2020)]{mccomas2020} McComas, D.~J., Bzowski, M., Dayeh, M. A., et al.\ 2020, \apjs, 248, 26
\bibitem[Malama et al. (2006)]{malama2006} Malama, Y. G., Izmodenov, V. V., Chalov, S. V. 2006, \aap, 445, 693-701
\bibitem[Richardson et al. (2008a)]{richardson2008a} Richardson, J. D., J. C. Kasper, C. Wang, J. W. Belcher, and A. J. Lazarus, 2008, Nature, 464, 63-66.
\bibitem[Richardson et al. (2008b)]{richardson2008b} Richardson, J. D. 2008, GeoRL, 35, L23104
\bibitem[Rucinski et al. (1993)]{rucinski1993} Rucinski, D., Fahr, H. J., and Grzedzielski, S. 1993, Planet. Space Sci. 41, 773
\bibitem[Scherer et al. (1998)]{scherer1998} Scherer, K., Fichtner, H., Fahr, H. J. 1998, J. Geophys. Res., 103, A2, 2105-2114
\bibitem[Schwadron et al. (2011)]{schwadron2011} Schwadron N. A., Allegrini F., Bzowski M. et al. 2011, \apj, 731, 56
\bibitem[Schwadron et al. (2014)]{schwadron2014} Schwadron N.A., Moebius E., Fuselier S.A. et al., 2014, \apjs, 215, 13
\bibitem[Shrestha et al. (2020)]{shrestha2020} Shrestha, B. L., Zirnstein, E. J., Heerikhuisen, J. 2020, \apj, 894, 102
\bibitem[Vasyliunas \& Siscoe (1976)]{vasyliunas1976} Vasyliunas, V.M., \& Siscoe, G.L. 1976, \jgr, 81, 7 
\bibitem[Witte(2004)]{witte2004} Witte, M.\ 2004, \aap, 426, 835 -- 84
\bibitem[Zank et al.(2010)]{zank2010} Zank, G. P., Heerikhuisen, J., Pogorelov, N. V., et al. 2010, \apj, 708, 1092
\bibitem[Zank et al.(2013)]{zank2013} Zank, G.~P., Heerikhuisen, J., Wood, B., et al.\ 2013, \apj, 763, 20 
%\bibitem[Zirnstein et al. (2013)]{zirnstein2013} Zirnstein E. J. 2013

%\bibitem[Zhao et al.(2019)]{zhao2019} Zhao, L.~L., Zank, G.~P., Adhikari, L.\ 2019, \apj, 879, 32

\bibitem[Zirnstein et al. (2014)]{zirnstein2014} Zirnstein, E. J., Heerikhuisen, J., Zank G. P. et al. 2014, \apj, 783, 129
\bibitem[Zirnstein et al. (2016)]{zirnstein2016} Zirnstein, E. J., Funsten, H. O., Heerikhuisen, J. et al. 2016, \apj, 826, 58
\bibitem[Zirnstein et al. (2017)]{zirnstein2017} Zirnstein, E. J., Heerikhuisen J.,  Zank G.P. et al. 2017, \apj, 836, 238
\end{thebibliography}

%\begin{thebibliography}{99}
%\end{thebibliography}

%%%%%%%%%%%%%%%%%%%%%%%%%%%%%%%%%%%%%%%%%%%%%%%%%%

%%%%%%%%%%%%%%%%% APPENDICES %%%%%%%%%%%%%%%%%%%%%

\appendix

\section{{\it IBEX-Hi} energy transmission} \label{app:flux_ibex-hi}

\begin{figure}
	\includegraphics[width=\columnwidth]{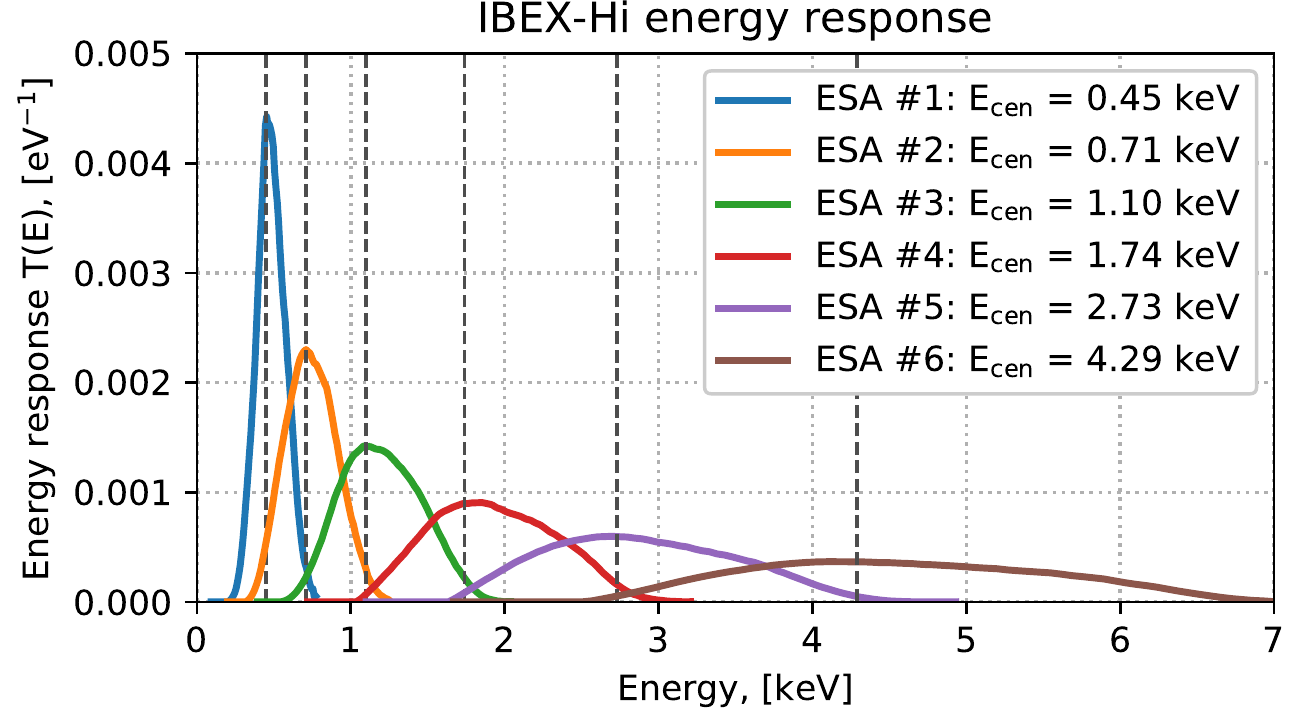}
	\caption{{\it IBEX-Hi} energy response as function of energy for all 6 energy channels. The calibration data files were taken from \url{http://ibex.swri.edu/ibexpublicdata/CalData/Hi/}.
	\label{fig:energy_resp}
	}
\end{figure}

To calculate the differential flux $J^{\rm M}_{\rm i}$ in the $\mathbf{LOS}$ direction measured by {\it IBEX-Hi} at the energy channel \#i, the energy transmission of the electrostatic analyzer (ESA) is taken into account:
\begin{equation}
J^{\rm M}_{\rm i}(\mathbf{LOS}) = \int^{E_{\rm i,max}}_{E_{\rm i.min}} j_{\rm ENA}(E, \mathbf{LOS}) T_{\rm i}(E) dE,\: {\rm i} = 1, ..., 6,
\label{eq:flux_energy_resp} 
\end{equation}
where superscript ``M" denotes the model, $j_{\rm ENA}(E, \mathbf{LOS})$ is differential spectra of ENA fluxes, $E_{\rm i, min}$ and $E_{\rm i, max}$ are the boundaries of ESA \#i accepting energies, $T_{i}(E)$ is normalized energy response function of the ESA \#i (see Figure \ref{fig:energy_resp}) such as $\int_{E_{\rm i, min}}^{E_{\rm i, max}} T_{\rm i}(E) dE = 1$. The point spread function of {\it IBEX-Hi} sensor is not taken into account in our modeling, so the fluxes are calculated for the center directions of the {\it IBEX} skymap $6^\circ \times 6^\circ$ bins.

\section{Model scaling factor} \label{app:scaling}
The difference between the model and data can be described in terms of the $\chi^2$ value (the weighted average of residuals):
\begin{equation}
	\chi^2(k) = \sum_{\rm ESA_i} \sum_{\rm LOS_j} \left( \frac{J^{\rm D}_{\rm i j} - k \times J^{\rm M}_{\rm i j}}{\sigma_{\rm i j}} \right)^2,
\label{eq:chi2}
\end{equation}
where $k$ is the scaling factor, $J^{\rm D}_{\rm i j}$ and $J^{\rm M}_{\rm i j}$ are the {\it IBEX-Hi} data and model ENA flux values (superscript ``D" denotes the data), $\sigma_{\rm i j}$ are the uncertainties of observations, the summations are performed for the top five {\it IBEX-Hi} energy channels ($i = 2,\ldots ,6$) and all $60 \times 30$ lines of sight for which the data is presented ($j = 1,\ldots , 1800$).

Using the weighted linear regression the best-fitting value $\hat k$, for which the $\chi^2(k)$ function takes its minimum, can be found as
\begin{equation}
	\hat k = \frac{\sum_{\rm i j} J^{\rm M}_{\rm i j} J^{\rm D}_{\rm i j} / \sigma_{\rm i j}^2 }{\sum_{\rm i j} (J^{\rm M}_{\rm i j} / \sigma_{\rm i j})^2},
\label{eq:scaling_factor}
\end{equation}
which is the solution of equation $d \chi^2/dk = 0$.

The reduced chi-square statistic $\chi^2_{\rm red}$, which is $\chi^2$ per degree of freedom, can be calculated as $\chi^2_{\rm red} = \chi^2 / \nu$, where $\nu = N - M$ equals the number of observations $N$ minus the number of fitted parameters $M$. In our study, the number of observations $N = 5 \times 1800 = 9000$ (5 energy channels and 1800 lines of sight).

%%%%%%%%%%%%%%%%%%%%%%%%%%%%%%%%%%%%%%%%%%%%%%%%%%

% Don't change these lines
\bsp	% typesetting comment
\label{lastpage}
\end{document}